\documentclass[a4paper]{article}

\usepackage{jheppub}                      
\usepackage{amsmath}
\usepackage{amssymb}   
\usepackage{mathrsfs}
\usepackage{braket}
\usepackage{slashed}
\usepackage{bm}
\usepackage{feynmp-auto}

\newcommand{\eq}[1]{eq.~\eqref{eq:#1}}
\newcommand{\eqs}[2]{eqs.~\eqref{eq:#1} and \eqref{eq:#2}}
\renewcommand{\sec}[1]{sec.~\ref{sec:#1}}

\newcommand{\fig}[1]{fig.~\ref{fig:#1}}

\newcommand{\sdt}{\!\cdot\!}
\newcommand{\nn}{\nonumber}

\newcommand{\tr}{\mathrm{tr}}
\newcommand{\df}{{\rm{d}}}

\def\img{{\rm i}}
\def\al{\alpha}

\def\ga{\gamma}

\def\de{\delta}

\def\eps{\epsilon}
\def\si{\sigma}

\newcommand{\thirteenbyninegrid}{
\fmfstraight
\fmfleft{a1,a2,a3,a4,a5,a6,a7,a8,a9}
\fmfright{m1,m2,m3,m4,m5,m6,m7,m8,m9}
\fmf{phantom}{a1,b1,c1,d1,e1,f1,g1,h1,i1,j1,k1,l1,m1}
\fmf{phantom}{a2,b2,c2,d2,e2,f2,g2,h2,i2,j2,k2,l2,m2}
\fmf{phantom}{a3,b3,c3,d3,e3,f3,g3,h3,i3,j3,k3,l3,m3}
\fmf{phantom}{a4,b4,c4,d4,e4,f4,g4,h4,i4,j4,k4,l4,m4}
\fmf{phantom}{a5,b5,c5,d5,e5,f5,g5,h5,i5,j5,k5,l5,m5}
\fmf{phantom}{a6,b6,c6,d6,e6,f6,g6,h6,i6,j6,k6,l6,m6}
\fmf{phantom}{a7,b7,c7,d7,e7,f7,g7,h7,i7,j7,k7,l7,m7}
\fmf{phantom}{a8,b8,c8,d8,e8,f8,g8,h8,i8,j8,k8,l8,m8}
\fmf{phantom}{a9,b9,c9,d9,e9,f9,g9,h9,i9,j9,k9,l9,m9}
}

\allowdisplaybreaks

\title{Transverse Momentum Measurements with Jets \\ at Next-to-Leading Power}

\author[a]{Rafael F. del Castillo,}
\author[b,c]{Max Jaarsma,}
\author[a]{Ignazio Scimemi,}
\author[b,c]{Wouter Waalewijn}

\affiliation[a]{Departamento de Física Teórica and IPARCOS, Facultad de Ciencias Físicas, Universidad Complutense Madrid, Plaza Ciencias 1, 28040 Madrid, Spain}
\affiliation[b]{ Institute for Theoretical Physics Amsterdam and Delta Institute for Theoretical Physics, University of
Amsterdam, Science Park 904, 1098 XH Amsterdam, The Netherlands}
\affiliation[c]{Nikhef, Theory Group, Science Park 105, 1098 XG, Amsterdam, The Netherlands }

\emailAdd{raffer06@ucm.es, m.jaarsma@uva.nl, ignazios@ucm.es, w.j.waalewijn@uva.nl}

\abstract{
In view of the increasing precision of theoretical calculations and experimental measurements, power corrections to transverse-momentum-dependent observables are highly important. We study the next-to-leading power corrections for transverse momentum measurements in $e^+e^- \to 2$ jets. We obtain a factorized expression for the cross section, which involve twist-2 and twist-3 operators, and identify the new jet functions that appear in it. We calculate these jet functions at order $\al_s$ for a family of recoil-free schemes, and provide the corresponding anomalous dimensions at leading order. Additionally, we show that the (endpoint) divergences that typically arise in sub-leading-power factorization can be subtracted and cancel for our case. By working with jets, everything is perturbatively calculable and there are substantial simplifications compared to the general next-to-leading power framework. Importantly, our analysis with jets can be extended to semi-inclusive deep-inelastic scattering, with the future Electron-Ion Collider as key application.
}

\preprint{IPARCOS-UCM-043}

\begin{document} 

\maketitle 

\section{Introduction}

The momentum distributions of quarks and gluons  (collectively called partons) inside hadrons is one of the most interesting subjects in Quantum Chromodynamics (QCD). These distributions are crucial for any prediction when hadrons are collided, such as the Large Hadron Collider (LHC) and the Electron Ion Collider (EIC). The fundamental theoretical tool for accessing these distributions is provided by factorization, which allows one to express (factorize) a cross section in terms of calculable coefficients and nonperturbative parton distributions. The universality of these distributions ensures that they can be measured in one process and then applied to another one. 

Parton distributions are correlation functions of partons  located at two different space-time points in a hadron. If these two points lie on  a light-cone direction, one gets the usual parton distribution functions (PDFs). When there is also a transverse separation, the correlation functions are transverse-momentum-dependent distributions  (TMDs). In principle one can also consider more general structures, such as Wigner distributions, generalized TMDs (GTMDs), etc., depending on more kinematic variables and involving different hadronic states, but they will not  be the subject of this work. Furthermore, one can also consider the different functions that account for the spin of partons or/and hadrons, so as to obtain the most general information about partonic entanglement~\cite{Boer:2004mv}. There are also corresponding nonperturbative distributions that describe (the fragmentation of) hadrons in the final state from energetic partons. We will focus on TMDs in the final state and work with jets rather than hadrons, because they can be described in perturbation theory. 

The factorization of the cross section for TMDs has a long history, starting with the seminal works of refs.~\cite{Collins:1981uk,Collins:1981va,Collins:1984kg}. It was observed that perturbative calculations involve a new type of divergences, nowadays called rapidity divergences. These divergences need special treatment and point to a more involved form of factorization than for the pure collinear case involving PDFs. TMD factorization has been achieved in more recent times, using special regulators for rapidity divergences and effective field theory~\cite{Collins:2011zzd,Echevarria:2011epo,Chiu:2012ir} or the background field method~\cite{Vladimirov:2021hdn}.
In these works, TMD factorization results from expanding the cross section in powers of $\lambda$, which is the ratio of a small transverse momentum $q_T$ and the scale $Q$ of the hard scattering process. In Drell-Yan (DY), semi-inclusive deep-inelastic scattering (SIDIS) and $e^+e^-\rightarrow 2$ jets (or 2 hadrons), $q_T$ is the small transverse momentum of the di-lepton  pair and $Q$ their invariant mass.

Restricting to the leading power in this expansion, and performing perturbative calculations for the renormalization group energy evolution, one has arrived at  the extraction of unpolarized TMDs up to next-to-next-to-next-to-next-to-leading logarithmic (N$^4$LL) order~\cite{Moos:2023yfa,Neumann:2022lft}. This is currently the highest logarithmic order that has been used to determine a nonperturbative distribution. The TMD formalism has also  been used to make a theoretical prediction for the $Z$-boson transverse momentum distribution at the same order~\cite{Camarda:2023dqn}.
In principle, the same accuracy can also be achieved  for the other, spin-dependent distributions described  by TMD factorization, because the TMD evolution kernel~\cite{Vladimirov:2016dll, Li:2016ctv, Vladimirov:2017ksc,Duhr:2022yyp,Moult:2022xzt}, cusp anomalous  dimension 
\cite{Moch:2004pa,Moch:2021qrk} and hard factors \cite{Kramer:1986sg,Matsuura:1988sm,Gehrmann:2010ue,Lee:2022nhh} don't depend on spin.

Given this high theoretical precision for the cross section at leading power, there have recently been advancements in the exploration of hadron structure at higher powers of $\lambda$. These power corrections are expected to introduce significant insights into our understanding of hadrons and their properties~\cite{Ebert:2018gsn,Vladimirov:2021hdn,Ebert:2021jhy,Rodini:2022wki,Gamberg:2022lju,Rodini:2023mnh,Rodini:2023plb}.
The need for power corrections is also clear from the per mille-level accuracy of recent DY measurements at the LHC.
Additionally, SIDIS and $e^+e^-$ experiments (like EIC~\cite{AbdulKhalek:2021gbh}, EIcC~\cite{Anderle:2021wcy} and Belle) are typically conducted at low energy, where power corrections are typically larger~\cite{Boussarie:2023izj}.  Summarizing, power corrections will enable higher precision in many experimental situations.  The process that we consider here can be important to establish the impact of NLP effects.

Recent progress has opened the possibility to study power corrections within the TMD formalism. In particular, we currently have at our disposal a complete basis of operators up next-to-leading power (NLP)~\cite{Vladimirov:2021hdn,Ebert:2021jhy}, whose evolution properties have been studied~\cite{Vladimirov:2021hdn,Rodini:2023mnh}. These studies have demonstrated that the factorization of the DY and SIDIS  processes remains valid up to NLP precision. However, additional non-perturbative operator matrix elements enter in the cross section, indicating the presence of new non-perturbative physics. Furthermore, the factorized formula for the cross section contains additional divergences characteristic of sub-leading-power factorization, which require careful treatment.

We find it desirable to have access to cross sections where non-perturbative QCD effects are minimal, to enable direct tests of NLP operators and shed light on the formalism. To this end, we investigate NLP effects using jets as hadronic final states. This is because jets are infrared-safe quantities whose properties can be largely fixed perturbatively, i.e.~they can be calculated in perturbation theory\footnote{Our jet functions may receive nonperturbative corrections suppressed by $\Lambda_{\rm QCD}^2/q_T^2$, which should not be confused with the $q_T/Q$ power corrections we consider here.} unlike the T-odd jets in refs.~\cite{Liu:2021ewb,Lai:2022aly} that require a  nonperturbative hypothesis. Specifically,  we are considering the impact of NLP corrections on di-jet production at $e^+e^-$ colliders. This offers the possibility to test NLP factorization at  $e^+e^-$ colliders such as LEP and Belle, and its extension to SIDIS will be particularly useful at the EIC~\cite{Abir:2023fpo}.

The possibility to use jets in TMD factorization has been explored in several works, see f.i.~\cite{Neill:2016vbi,Kang:2017glf,Liu:2018trl,Gutierrez-Reyes:2018qez,Gutierrez-Reyes:2019vbx,Arratia:2019vju,Chien:2020hzh,Arratia:2020nxw,Liu:2020dct,Kang:2021ffh,Chien:2022wiq,Arratia:2022oxd}. Refs.~\cite{Gutierrez-Reyes:2018qez,Gutierrez-Reyes:2019vbx} focused mainly on the consistency of the TMD factorization theorems with jet definitions and algorithms, finding that the definition of the jet axis and radius are essential to establish the factorization. The standard jet definition suffers from non-global logarithms, due to its sensitivity to whether soft radiation gets clustered into the jet or not, substantially limiting the accuracy. By contrast, recoil-insensitive schemes~\cite{Larkoski:2014uqa} such as the Winner-Takes-All (WTA) jet axis~\cite{Bertolini:2013iqa} are insensitive to this effect, enabling high precision calculations.

In this work, we address two fundamental questions: First, we want to identify the new jet functions that arise from the additional operators at NLP. Second, we want to obtain the factorized expression for the cross section in terms of jet functions and hard matching coefficients. The answers to these questions that we obtain, enable us to study how NLP corrections impact transverse-momentum-dependent measurements, including angular asymmetries. We find that in order to get a non-vanishing contribution from NLP operators, one has to break the symmetry between the energetic radiation going into the two different directions. For the case of di-jet production,  this can be achieved by e.g.~using two different (recoil-free) recombination schemes.  

Despite the  complexity of the subject we have made an effort to present our material in a didactic way. The choice of the process and the jet definitions that we use, greatly reduces the number of independent distributions and simplifies the evolution with respect to the original papers~\cite{Braun:2009mi,Vladimirov:2021hdn,Rodini:2022wic,Rodini:2023plb}, see \eq{ga_J_21}. In this sense, the present work can also be viewed as an introductory overview to the study of NLP effects in TMD factorization. 

The outline of this paper is as follows: We start in sec.~\ref{sec:Kinematics} by introducing the kinematic variables needed to describe $e^+e^-\rightarrow 2$ jets, and writing the cross section in terms of the hadronic and leptonic tensor.  In sec.~\ref{sec:HadronicTensor},  the hadronic tensor is factorized and rewritten, with Fierz identities relegated to app.~\ref{sec:Fierz}. Our final expression for the cross section in terms of jet functions is given in \eq{sigmabare}. In sec.~\ref{sec:Divergences}, we discuss the cancellation of rapidity divergences and additional divergences that show up at sub-leading-power. This section also covers the renormalization and evolution of the jet functions. We calculate and present all ingredients of the cross section to first order in $a_s$ in sec.~\ref{sec:JetFunction}. Our conclusions are given in sec.~\ref{sec:Conclusions}.

\section{Kinematics}
\label{sec:Kinematics}

The process that we consider in this paper is the annihilation of an electron and positron, with momenta $\ell$ and $\ell'$, into two jets with momenta $P_1$ and $P_2$,
\begin{align}
    e^-(\ell)\,e^+(\ell')\to j_1(P_1)\,j_2(P_2)\, X\,.
\end{align}
We will assume a large jet radius, such that all collinear final-state particles are clustered in one of the two jets, and any radiation outside the jets $X$ is soft\footnote{To be precise, we assume $Q R \gg q_T$, such that the transverse momentum measurement restricts the collinear radiation to be inside the jets.}. At lowest order in the electromagnetic coupling, this process proceeds by the electron and positron annihilating into an intermediate photon with momentum $q=\ell+\ell'$, which subsequently creates a quark-antiquark pair that produces two jets. In the following, we assume that the jet algorithm yields massless jet momenta,  $P_1^2=P_2^2=0$ (as in the WTA recombination scheme), and we introduce light-like vectors $n$ and $\bar n$  such that 
\begin{align}
    P_1^\mu=P_1^- n^\mu, \quad P_2^\mu=P_2^+ \bar n^\mu, \quad \textrm{and}
    \quad n\cdot \bar n=1.
\end{align}

\begin{figure}[t]
\centering
\includegraphics[width=0.6\textwidth]{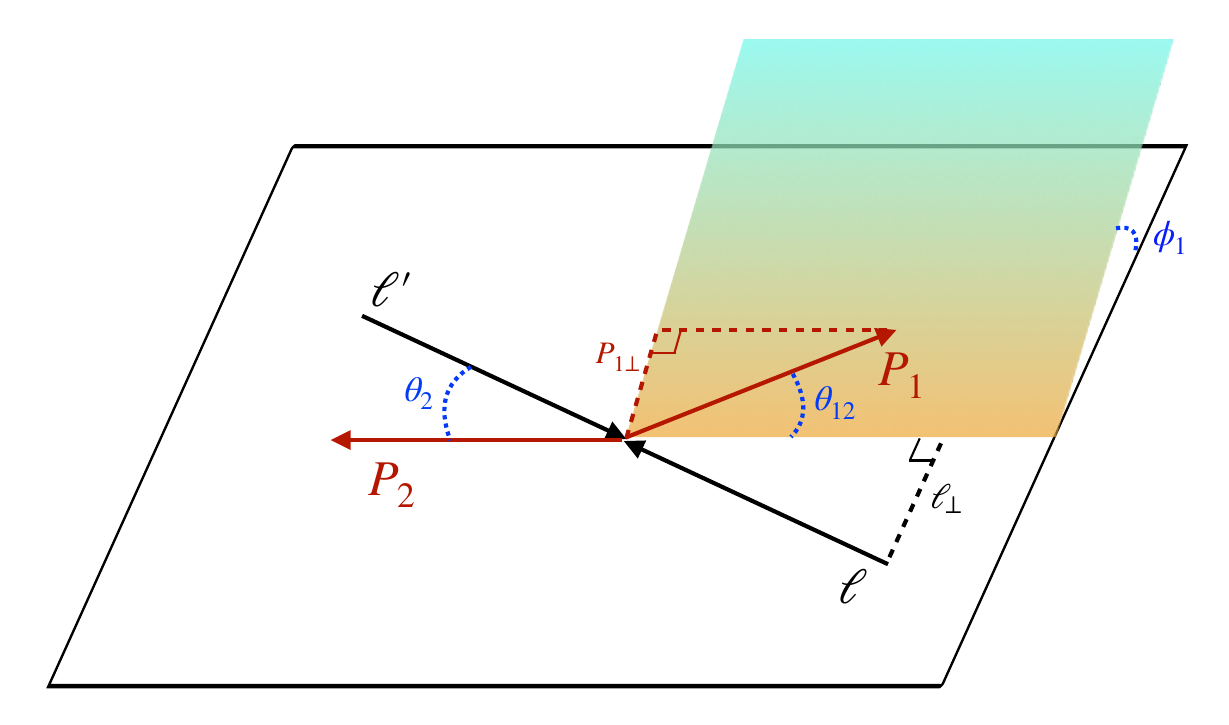}
\caption{\label{fig:epluseminus} Kinematics of the di-jet production in $e^+e^-$.}
\end{figure}

In order to introduce our observables, we define two planes, which we will refer to as the transverse and perpendicular plane. The transverse plane is orthogonal to the two outgoing jets and the perpendicular plane is orthogonal to the intermediate photon momentum $q$ and one of the outgoing jets. These planes can be described in terms of the metric tensors that project onto the corresponding sub-spaces, 
\begin{align}\label{eq:gmnT}
    g_T^{\mu\nu}
    &=
    g^{\mu\nu}
    -
    \frac{1}{P_1\sdt P_2}
    \bigl(P_1^\mu P_2^\nu + P_2^\mu P_1^\nu\bigr),
    \\
    g_{\perp}^{\mu\nu}
    &=
    g^{\mu\nu}
    +
    \frac{Q^2}{(P_2 \sdt q)^2} P_2^\mu P_2^\nu
    -\frac{1}{P_2 \sdt q}\bigl(q^\mu P_2^\nu+P_2^\mu q^\nu\bigr),
\end{align}
and we define the transverse and perpendicular components of a vector $v$ as
\begin{align}
    v_T^\mu=g_T^{\mu\nu}v_\nu\,,\qquad
    v_\perp^\mu=g_\perp^{\mu\nu}v_\nu\,.
\end{align}
Note that, by construction,
\begin{align}
    P_{1,T}^\mu = P_{2,T}^\mu=0\,,\qquad
    q_\perp^\mu=P_{2,\perp}^\mu=0\,.
\end{align}
We will sometimes write two-dimensional vectors in boldface notation, where it is implied that the two-dimensional metric has positive signature. As we mostly encounter transverse vectors, we will suppress the subscript $T$, e.g.~${\bf b} = {\bf b}_T$.

We now list the variables that we use to describe the cross section, which are the same as in ref.~\cite{Boer:1997mf} for the massless case:
\begin{align} \label{eq:obs}
    z_i=\frac{P_i\cdot q}{\ell\cdot q}\,,\qquad
    y=\frac{P_2\cdot \ell}{P_2\cdot q}\,,\qquad
        Q_T^2 = - q_T^2 = -g_T^{\mu\nu} q_\mu q_\nu\,, \qquad
            Q^2=q^2\,,
\end{align}
as well as an azimuthal angle 
\begin{align}
    \cos \phi_1 =\frac{\boldsymbol{\ell}_\perp\cdot\mathbf{P}_{1\perp}}
        {|\boldsymbol{\ell}_\perp||\mathbf{P}_{1\perp}|}
    \qquad \text{or} \qquad
    \cos\alpha
    &=
    \frac{\boldsymbol{\ell}\cdot\mathbf{q}}{|\boldsymbol{\ell}||\mathbf{q}|}    
\,.\end{align}
The angles $\phi_1$ and $\alpha$ are not independent. It   is natural to use the former in the center-of-mass frame and as depicted in~\fig{epluseminus}, while it is convenient to use the latter in the frame where the jets are exactly back-to-back. They are related by
\begin{align}
    \cos\phi_1&=
    \pm\cos\alpha\sqrt{1
        -\frac{Q_T^2}{Q^2}\frac{y}{1-y}\sin^2\alpha}
    +\frac{Q_T}{Q}\sqrt{\frac{y}{1-y}}\sin^2\alpha
    =\cos\alpha+\mathcal{O}\Bigl(\frac{Q_T}{Q}\Bigr)\,.
\end{align}
All dot products of four-vectors can be expressed in terms of the observables listed above.

 While the above expressions are Lorentz invariant, the interpretation we now give is valid for the (di-lepton) center-of-mass frame: First of all, the $z_i$ ($i=1,\;2$)  are the energy fractions of the two jets. The variable $y$ corresponds to the angle $\theta_2$ between the momentum of the jet and the lepton, $y = (1 -  \cos \theta_2)/2$. Next, $\phi_1$ is the azimuthal angle between the perpendicular component of jet 1 and the electron momentum $\ell$. Finally, 
\begin{align}
  Q_T^2 = - Q^2 + \frac{z_1 z_2 Q^4}{2 P_1 \sdt P_2}  \approx \tfrac14 \theta_{12}^2 Q^2
\,,\end{align}
where $\theta_{12}$ is the angular decorrelation of the two jets. These angles are shown in~\fig{epluseminus}.
Note that the invariant mass $Q$ of the collision is not an observable, but it is determined by the initial state of the experiment. 

In this paper we consider the large jet radius limit, where $z_1=z_2=1$ up to power corrections, so we will not be differential in them. This leads us to consider the following cross section 
\begin{align}
    \frac{\df\sigma}{\df y\, \df^2 {\bf q}}
    &=
    \frac{\pi \alpha_{\rm em}^2Q_q^2 }{Q^4}
    L_{\mu\nu} W^{\mu\nu}\,,
\end{align}
where $L_{\mu\nu}$  and $W^{\mu\nu}$ are respectively the leptonic and hadronic tensor. The leptonic tensor reads
\begin{align} \label{eq:leptonic}
    L_{\mu\nu}&=2\bigl(\ell_\mu\ell'_\nu+\ell'_\mu\ell_\nu\bigr)-Q^2 g_{\mu\nu}+2 \img\, \lambda_e\,\epsilon_{\mu\nu\rho\sigma} \ell^\rho\ell^{\prime\sigma}\,,\quad\epsilon_{0123}=1,
\end{align}
and in this work we will neglect the lepton helicity contribution proportional to $\lambda_e$. Our main focus is thus on the hadronic tensor, which contains all QCD effects. For this, the recent work of ref.~\cite{Vladimirov:2021hdn} opens the door to a more systematic study of the hadronic tensor up to NLP,  and this is the subject of the next section.

The case of two hadrons (instead of jets) 
has been studied before, especially in relation to Belle experiment, see the reviews in refs.~\cite{Boussarie:2023izj,USBelleIIGroup:2022qro,Accardi:2022oog,Belle:2008fdv}. The distributions in this case were classified in ref.~\cite{Boer:1997mf}, and mass corrections to kinematics have been included in ref.~\cite{Mulders:2019mqo}.

\section{Factorization at NLP}
\label{sec:HadronicTensor}

The goal of this section is to obtain a factorized expression for the cross section in terms of a set of jet functions and hard functions. The expression we obtain holds at the bare and unsubtracted level, and we postpone the discussion of renormalization and (overlap) subtraction for the next section.

To obtain a factorized expression for the cross section we first expand the hadronic tensor up to next-to-leading power using the TMD operator product expansion method. The complete and un-expanded hadronic tensor for the process under consideration is given by,
\begin{align} \label{eq:had_tensor}
    W^{\mu\nu}(q)&=\int\frac{\df^4 y}{(2\pi)^4}\,e^{\img q\cdot y}\, \sum_X
    \bra{0}J^{\mu\,\dagger}(y)\ket{J^n_\text{alg} J^{\bar{n}}_\text{alg} X}
    \bra{J^n_\text{alg} J^{\bar{n}}_\text{alg} X}J^\nu(0)\ket{0}\,,
\end{align}
where $J^\mu(y)$ is the electromagnetic current
\begin{align} \label{eq:em_current}
    J^\mu(y)=\bar{\psi}(y)\gamma^\mu\psi(y)\,.
\end{align}
Since we assume a large jet radius, all energetic final-state particles are clustered into one of the two jets and $X$ only contains soft radiation. The explicit notation that we use for the  power expansion of the hadronic tensor is
\begin{align}
    W^{\mu\nu}(q)&=W^{\mu\nu}_\text{LP}(q)+W^{\mu\nu}_\text{NLP}(q)
    +\mathcal{O}\Bigl(\frac{1}{Q^2}\Bigr)\,.
\end{align}

In order to arrive at our final expressions for $W_{\text{LP},\text{NLP}}$, we proceed in several steps, starting from the TMD operator expansion method and the basis of operators in sec.~\ref{sec:OB}. In particular, we list a set of  operator building blocks that have definite twist, which allow for a clean separation in terms of power counting. We use these definitions in sec.~\ref{sec:EHT}, to obtain the power expansion of the product of currents in \eq{had_tensor} in position space (indicated by the tilde)
\begin{align}
    \tilde{\mathcal{W}}^{\mu\nu}(y)&=J^{\mu \dagger}(y)\, J^\nu(0)
    =\tilde{\mathcal{W}}^{\mu\nu}_\text{LP}(y)+\tilde{\mathcal{W}}^{\mu\nu}_\text{NLP}(y)\,.
    \label{eq:WJJ}
\end{align}
Next, we include the external jet states and Fourier transform back to momentum space, to obtain an expression for the hadronic tensor in terms of a set of jet functions with open spinor indices. In sec.~\ref{sec:JF}, we apply Fierz relations to obtain a set of jet functions with closed spinor indices, and use discrete symmetries and Lorentz decomposition to simplify and reduce the basis of jet functions, resulting in just one new jet function at NLP! Sec.~\ref{sec:factorization} contains the final expression for the hadronic tensor in terms of jet functions, as well as its contraction with the leptonic tensor.

\subsection{Operator basis}
\label{sec:OB}

In this work we make use of the TMD operator product expansion \cite{Vladimirov:2021hdn}. In this method, background fields are introduced for the two collinear directions
\begin{align}
    \psi(x)=\chi(x) + q_n(x) + q_{\bar{n}}(x)\,,\qquad
    A^\mu(x)=B^\mu(x) + A^\mu_n(x) + A^\mu_{\bar{n}}(x)\,.
\end{align}
The background fields $q_n, q_{\bar{n}}, A^\mu_n, A^\mu_{\bar{n}}$ will be used to construct our operators, while the fields $\chi, B^\mu$ are purely dynamical. As usual, we decompose the quark background fields into ``good" and ``bad" components,
\begin{align}
    q_{\bar{n}}(x)=\xi_{\bar{n}}(x)+\eta_{\bar{n}}(x)\,,\qquad
        q_{n}(x)=\xi_{n}(x)+\eta_{n}(x)\,,
\end{align}
where
\begin{align} \label{eq:good_bad}
\xi_{\bar n}(x)&=\frac{\slashed{\bar n}\slashed{n}}{2}q_{\bar n}(x),& \eta_{\bar n}(x)&=\frac{\slashed{n}\slashed{\bar n}}{2}q_{\bar n}(x),
\nn \\
\xi_{n}(x)&=\frac{\slashed{n}\slashed{\bar n}}{2}q_{n}(x),& \eta_{n}(x)&=\frac{\slashed{\bar n}\slashed{n}}{2}q_{n}(x)\,,
\end{align}
such that each field has a definite power counting ($\xi\sim\lambda$, $\eta\sim \lambda^2$). The collinear and anti-collinear regions overlap in the soft region. To avoid the double counting of this region its contribution must be carefully subtracted. We discuss the subtraction of the overlap region in sec.~\ref{sec:overlap}.

Using the background fields one can construct a set of building-block operators of definite twist (dimension minus spin), 
\begin{align} \label{eq:U1}
    U_{1,\bar{n}}(y^-,b)
    &=
    [Ln+b,y^-n+b]\,\xi_{\bar{n}}(y^- n + b)\,,
    \nn \\
    U_{1,n}(y^+,b)
    &=
    [\bar{L}\bar{n}+b,y^+\bar{n}+b]\,\xi_n(y^+ \bar{n} + b)\,,
    \\[1ex] \label{eq:U2}
    U_{2,\bar{n}}(\{y_1^-,y_2^-\},b)
    &=
    g\,[L n + b,y_2^- n + b]\,
    \gamma_{T,\rho} F_{\bar n}^{\rho+}(y_2^- n + b) 
    [y_2^- n + b,y_1^- n + b]\,\xi_{\bar{n}}(y_1^- n + b)\,,
    \nn \\
    U_{2,n}(\{y_1^+,y_2^+\},b)
    &=
    g\,[\bar{L} \bar{n} + b,y_2^+ \bar{n} + b]\,
    \gamma_{T,\rho} F_{n}^{\rho-}(y_2^+ \bar{n} + b) 
    [y_2^+ \bar{n} + b,y_1^+ \bar{n} + b]\,\xi_n(y_1^+ \bar n + b)\,.
\end{align}
Here, the first subscript on the operator denotes the twist of the operator and the second subscript denotes the collinear sector to which the operator belongs. These operators contain Wilson lines, which we denote by
\begin{align}
    [a,b]=P\exp\biggl[-\img g\int_a^b \df z_\mu\,A^\mu(z)\biggr]\,.
\end{align}
The gauge field appearing in the Wilson line matches the collinear direction under consideration, e.g.~$A_n$ ($A_{\bar n}$) for the Wilson line in  $U_{2,n}$ ($U_{2,\bar n}$).
The operators in \eqs{U1}{U2} will enter in the matrix-element definition of the jet functions. Note that these operators are process dependent via their dependence on $L$ and $\bar{L}$, with $L=-\infty$ for incoming partons and $L=+\infty$ for outgoing partons. Since we consider jet production, we set $L=\bar{L} = +\infty$ from now on. In the Wilson lines we will also introduce  a $\delta$-regulator whenever necessary  (see refs.~\cite{Vladimirov:2021hdn,Echevarria:2016scs} for more details).

\subsubsection*{Operator basis at LP}

At lowest order in the power expansion of the product of currents in \eq{WJJ}, for which the result will be shown in sec.~\ref{sec:EHT}, the following combinations of operators appear
\begin{align}
    \mathcal{O}_{11,\bar{n}}^{ij}(\{y^-,0\},y_T)
    &=
    \bar{\xi}_{\bar{n},j}^{(-)}(y^- n + y_T)\,
    \xi_{\bar{n},i}^{(+)}(0)\,,
    \nn \\
    \mathcal{O}_{11,n}^{ij}(\{y^+,0\},y_T)
    &=
    \bar{\xi}_{n,j}^{(-)}(y^+ \bar{n} + y_T)\,
    \xi_{n,i}^{(+)}(0)\,,
    \nn \\[2ex]
    \overline{\mathcal{O}}_{11,\bar{n}}^{ij}(\{y^-,0\},y_T)
    &=
    \tr_c\bigl[\xi_{\bar{n},i}^{(-)}(y^- n + y_T)\,
    \bar{\xi}_{\bar{n},j}^{(+)}(0)\bigr]\,,
    \nn \\
    \overline{\mathcal{O}}_{11,n}^{ij}(\{y^+,0\},y_T)
    &=
    \tr_c\bigl[ \xi_{n,i}^{(-)}(y^+ \bar{n} + y_T)\,
    \bar{\xi}_{n,j}^{(+)}(0)\bigr]\,.
\end{align}
The $(\pm)$ superscript denote causal and anti-causal fields from the Keldysh formalism~\cite{Keldysh:1964ud}.
The color indices of the fields are contracted, which is made explicit for $\overline{\mathcal{O}}_{11}$ by the color trace $\tr_c$, and $i,j$ are spin indices.
The gauge invariance in the expressions above is restored using
the building-block operators in  \eqs{U1}{U2}, by
\begin{align} \label{eq:O11}
    \mathcal{O}_{11,\bar{n}}^{ij}
    &=
    \bigl[\overline{U}_{1,\bar{n}}(y^-,y_T)\bigr]_j^{(-)}
    \bigl[U_{1,\bar{n}}(0,0)\bigr]_i^{(+)}\,,
    \nn \\
    \mathcal{O}_{11,n}^{ij}
    &=
    \bigl[\overline{U}_{1,n}(y^+,y_T)\bigr]_j^{(-)}
    \bigl[U_{1,n}(0,0)\bigr]_i^{(+)}\,,
    \nn \\[2ex]
    \overline{\mathcal{O}}_{11,\bar{n}}^{ij}
    &=
    \tr_c\bigl\{ \bigl[U_{1,\bar{n}}(y^-,y_T)\bigr]_i^{(-)}
    \bigl[\overline{U}_{1,\bar{n}}(0,0)\bigr]_j^{(+)}\bigr\}\,,
    \nn \\
    \overline{\mathcal{O}}_{11,n}^{ij}
    &=
        \tr_c\bigl\{ \bigl[U_{1,n}(y^+,y_T)\bigr]_i^{(-)}
    \bigl[\overline{U}_{1,n}(0,0)\bigr]_j^{(+)}\bigr\}\,.
\end{align}

\subsubsection*{Operator basis at NLP}

At next-to-leading power in the expansion of \eq{WJJ}, one finds contributions from the following operators
\begin{align} 
    \mathbb{O}_{21,\bar{n}}^{ij}(\{y^-,y^-,0\},y_T)
    &=
    \bigl[\bar{\xi}_{\bar{n}}^{(-)} 
        \slashed{A}_{\bar{n},T}^{(-)}(y^- n + y_T)\bigr]_j\,
    \xi_{\bar{n},i}^{(+)}(0)\,,
    \nn \\
    \mathbb{O}_{21,n}^{ij}(\{y^+,y^+,0\},y_T)
    &=
    \bigl[\bar{\xi}_{n}^{(-)} 
        \slashed{A}_{n,T}^{(-)}(y^+ \bar{n} + y_T)\bigr]_j\,
    \xi_{n,i}^{(+)}(0)\,,
    \nn \\[2ex]
    \overline{\mathbb{O}}_{21,\bar{n}}^{ij}(\{y^-,y^-,0\},y_T)
    &=
    \tr_c\bigl\{\bigl[\slashed{A}_{\bar{n},T}^{(-)}
        \xi_{\bar{n}}^{(-)}(y^- n + y_T)\bigr]_i\,
    \bar{\xi}_{\bar{n},j}^{(+)}(0)\bigr\}\,,
    \nn \\
    \overline{\mathbb{O}}_{21,n}^{ij}(\{y^+,y^+,0\},y_T)
    &=
    \tr_c\bigl\{\bigl[\slashed{A}_{n,T}^{(-)}
        \xi_{n}^{(-)}(y^+ \bar{n} + y_T)\bigr]_i\,
    \bar{\xi}_{n,j}^{(+)}(0)\bigr\}\,,
    \nn \\[2ex]
    \mathbb{O}_{12,\bar{n}}^{ij}(\{y^-,0,0\},y_T)
    &=
    \bar{\xi}_{\bar{n},j}^{(-)}(y^- n + y_T)\,
    \bigl[\slashed{A}_{\bar{n},T}^{(+)} \xi_{\bar{n}}^{(+)}(0)\bigr]_i\,,
    \nn \\
    \mathbb{O}_{12,n}^{ij}(\{y^+,0,0\},y_T)
    &=
    \bar{\xi}_{n,j}^{(-)}(y^+ \bar{n} + y_T)\,
    \bigl[\slashed{A}_{n,T}^{(+)} \xi_{n}^{(+)}(0)\bigr]_i\,,
    \nn \\[2ex]
    \overline{\mathbb{O}}_{12,\bar{n}}^{ij}(\{y^-,0,0\},y_T)
    &=
    \tr_c\bigl\{\xi_{\bar{n},i}^{(-)}(y^- n + y_T)\,
    \bigl[\bar{\xi}_{\bar{n}}^{(+)} \slashed{A}_{\bar{n},T}^{(+)}(0)\bigr]_j\bigr\}\,,
    \nn \\
    \overline{\mathbb{O}}_{12,n}^{ij}(\{y^+,0,0\},y_T)
    &=
    \tr_c\bigl\{\xi_{n,i}^{(-)}(y^+ \bar{n} + y_T)
    \bigl[\bar{\xi}_{n}^{(+)} \slashed{A}_{n,T}^{(+)}(0)\bigr]_j\bigr\}\,.
    \label{eq:O12}
\end{align}
To write these in terms of the building-block operators we need to do a few manipulations. First we replace the transverse gauge fields that appear here by a field strength via the following relations, 
\begin{align}
    A^\mu_{\bar{n}}(y)
    =
    g\int_0^{\infty}\df z^-\,e^{-z^- \delta^+}\,
    F^{\mu+}_{\bar{n}}(y + z^- n)\,,
    \qquad
    A^\mu_{n}(y)
    =
    g\int_0^{\infty}\df z^+\,e^{-z^+ \delta^-}\,
    F^{\mu-}_{n}(y + z^+ \bar{n})\,,
\end{align}
which introduces a $\delta$-regulator for rapidity divergences.
Subsequently, we apply the following identity 
\begin{align}\nn
    \int_0^\infty\df z\,e^{-z\delta}\,
    f(z)
    &=
    -\img\int_{-\infty}^{\infty}\df \xi\,
    \frac{1}{\pm\xi - \img\delta_E^\pm}
    \int\frac{\df z}{2\pi}\,e^{\pm \img \xi q^\pm z}\,
    f(z)\,,
\end{align}
where a positive $\delta_E^\pm=\delta^\pm/q^\pm$ ensures that the integral is well defined and shows how $\delta$ regulates rapidity divergences. Note that in this identity the sign of $\xi$ may be chosen freely as it is integrated over all values. It turns out that it is convenient to choose $+\xi$ for the $\mathbb{O}_{21}$ operators and $-\xi$ for the $\mathbb{O}_{12}$ operators, as the jet functions that will be built from these operators then have support for $\xi\in[0,1]$. This convention differs from earlier work, and allows for several simplifications in our case.

The above manipulations allow us to express the operators in \eq{O12}  in the NLP hadronic tensor in terms of the building-block operators in \eqs{U1}{U2} as follows,
\begin{align}\nn
    \mathbb{O}_{21,\bar{n}}^{ij}
    &=
    -\img\int\df\xi\,\frac{1}{\xi - \img\delta_E^+}
    \int\frac{\df z^-}{2\pi}\,e^{\img \xi (z^- - y^-) q^+}\,
    \bigl[\overline{U}_{2,\bar{n}}(\{y^-,z^-\},y_T)\bigr]_j^{(-)}
    \bigl[U_{1,\bar{n}}(0,0)\bigr]_i^{(+)}\,,
    \\ \nn
    \mathbb{O}_{21,n}^{ij}
    &=
    -\img\int\df\xi\,\frac{1}{\xi - \img\delta_E^-}
    \int\frac{\df z^+}{2\pi}\,e^{\img \xi (z^+ - y^+) q^-}\,
    \bigl[\overline{U}_{2,n}(\{y^+,z^+\},y_T)\bigr]_j^{(-)}
    \bigl[U_{1,n}(0,0)\bigr]_i^{(+)}\,,
    \\[2ex] \nn
    \overline{\mathbb{O}}_{21,\bar{n}}^{ij}
    &=
    -\img\int\df\xi\,\frac{1}{\xi - \img\delta_E^+}
    \int\frac{\df z^-}{2\pi}\,e^{\img \xi (z^- - y^-) q^+}\,
    \tr_c\bigl\{ \bigl[U_{2,\bar{n}}(\{y^-,z^-\},y_T)\bigr]^{(-)}_i
    \bigl[\overline{U}_{1,\bar{n}}(0,0)\bigr]_j^{(+)} \bigr\} \,,
    \\ \nn
    \overline{\mathbb{O}}_{21,n}^{ij}
    &=
    -\img\int\df\xi\,\frac{1}{\xi - \img\delta_E^-}
    \int\frac{\df z^+}{2\pi}\,e^{\img \xi (z^+ - y^+) q^-}\,
    \tr_c\bigl\{ \bigl[U_{2,n}(\{y^+,z^+\},y_T)\bigr]^{(-)}_i
    \bigl[\overline{U}_{1,n}(0,0)\bigr]_j^{(+)} \bigr\}\,,
    \\[2ex] \nn
    \mathbb{O}_{12,\bar{n}}^{ij}
    &=
    \img\int\df\xi\,\frac{1}{\xi + \img\delta_E^+}
    \int\frac{\df z^-}{2\pi}\,e^{-\img \xi z^- q^+}\,
    \bigl[\overline{U}_{1,\bar{n}}(y^-,y_T)\bigr]_j^{(-)}
    \bigl[U_{2,\bar{n}}(\{0,z^-\},0)\bigr]^{(+)}_i\,,
    \\ \nn
    \mathbb{O}_{12,n}^{ij}
    &=
    \img\int\df\xi\,\frac{1}{\xi + \img\delta_E^-}
    \int\frac{\df z^+}{2\pi}\,e^{-\img \xi z^+ q^-}\,
    \bigl[\overline{U}_{1,n}(y^+,y_T)\bigr]_j^{(-)}
    \bigl[U_{2,n}(\{0,z^+\},0)\bigr]^{(+)}_i\,,
    \\[2ex] \nn
    \overline{\mathbb{O}}_{12,\bar{n}}^{ij}
    &=
    \img\int\df\xi\,\frac{1}{\xi + \img\delta_E^+}
    \int\frac{\df z^-}{2\pi}\,e^{-\img \xi z^- q^+}\,
    \tr_c\bigl\{ \bigl[U_{1,\bar{n}}(y^-,y_T)\bigr]_i^{(+)}
    \bigl[\overline{U}_{2,\bar{n}}(\{0,z^-\},0)\bigr]_j^{(+)} \bigr\}\,,
    \\ 
    \overline{\mathbb{O}}_{12,n}^{ij}
    &=
    \img\int\df\xi\,\frac{1}{\xi + \img\delta_E^-}
    \int\frac{\df z^+}{2\pi}\,e^{-\img \xi z^+ q^-}\,
    \tr_c\bigl\{ \bigl[U_{1,n}(y^+,y_T)\bigr]_i^{(+)}
    \bigl[\overline{U}_{2,n}(\{0,z^+\},0)\bigr]_j^{(+)} \bigr\}\,.
    \label{eq:OOs}
\end{align}
We have omitted the space-time arguments on the operators on the left-hand side as they are the same as in eq.~(\ref{eq:O12}).
The sign in front of $\img\delta$ differs between $\mathbb{O}_{12}$ and $\mathbb{O}_{21}$. As we will see, their matrix elements are related by complex conjugation.

\subsection{Expansion of the hadronic tensor}
\label{sec:EHT}

The aim of this section is to write the hadronic tensor in terms of matrix elements of the building-block operators in sec.~\ref{sec:OB}. For this, we first consider the power expansion of the product of currents in \eq{WJJ} in position space. Following \cite{Vladimirov:2021hdn}, the leading power contribution  can be written  at leading perturbative order as
\begin{align}
    \tilde{\mathcal{W}}_{\text{LP}}^{\mu\nu}(y)
    &=
    \frac{1}{N_c} (\gamma_T^\mu)_{ij} (\gamma_T^\nu)_{kl}
    \Bigl(
    \mathcal{O}_{11,\bar{n}}^{li}\,
    \overline{\mathcal{O}}_{11,n}^{jk}
    +
    \overline{\mathcal{O}}_{11,\bar{n}}^{jk}\,
    \mathcal{O}_{11,n}^{li}
    \Bigr)\,.
    \label{eq:WLP}
\end{align}
For brevity we suppressed the space-time arguments of the operators, but they are the same as in \eq{O11}.
At next-to-leading power more operators appear and the expression for the hadronic tensor is a bit more involved, 
\begin{align}
    \tilde{\mathcal{W}}_{\text{NLP}}^{\mu\nu}(y)
    &=
    -\frac{1}{N_c}
    \bigl[n^\mu (\gamma_T^\rho)_{ij} (\gamma_T^\nu)_{kl} 
        + n^\nu (\gamma_T^\mu)_{ij} (\gamma_T^\rho)_{kl}\bigr]
    \biggl[
    \biggl(\frac{\partial_\rho}{\partial_+}
    \mathcal{O}_{11,\bar{n}}^{li}\biggr)\,
    \overline{\mathcal{O}}_{11,n}^{jk}
    +
    \biggl(\frac{\partial_\rho}{\partial_+}
    \overline{\mathcal{O}}_{11,\bar{n}}^{jk}\biggr)\,
    \mathcal{O}_{11,n}^{li}
    \biggr]
    \nonumber
    \\
    &\quad
    -
    \frac{1}{N_c}
    \bigl[\bar{n}^\mu (\gamma_T^\rho)_{ij} (\gamma_T^\nu)_{kl} 
        + \bar{n}^\nu (\gamma_T^\mu)_{ij} (\gamma_T^\rho)_{kl}\bigr]
    \biggl[
    \mathcal{O}_{11,\bar{n}}^{li}\,
    \biggl(\frac{\partial_\rho}{\partial_-}
    \overline{\mathcal{O}}_{11,n}^{jk}\biggr)
    +
    \overline{\mathcal{O}}_{11,\bar{n}}^{jk}\,
    \biggl(\frac{\partial_\rho}{\partial_-}
    \mathcal{O}_{11,n}^{li}\biggr)
    \biggr]
    \nonumber
    \\
    &\quad
    +\frac{\img g}{N_c}\,
    \delta_{ij} (\gamma_T^\nu)_{kl}
    \biggl[
    \mathbb{O}_{21,\bar{n}}^{li}
    \biggl(\frac{\bar{n}^\mu}{\overrightarrow{\partial_-}}
        -\frac{n^\mu}{\overleftarrow{\partial_+}}\biggr)
    \overline{\mathcal{O}}_{11,n}^{jk}
    -
    \overline{\mathbb{O}}_{21,\bar{n}}^{jk}
    \biggl(\frac{\bar{n}^\mu}{\overrightarrow{\partial_-}}
        -\frac{n^\mu}{\overleftarrow{\partial_+}}\biggr)
    \mathcal{O}_{11,n}^{li}
    \nonumber
    \\
    &\quad\qquad\quad
    +
    \mathcal{O}_{11,\bar{n}}^{li}
    \biggl(\frac{\bar{n}^\mu}{\overrightarrow{\partial_-}}
        -\frac{n^\mu}{\overleftarrow{\partial_+}}\biggr)
    \overline{\mathbb{O}}_{21,n}^{jk}
    -
    \overline{\mathcal{O}}_{11,\bar{n}}^{jk}
    \biggl(\frac{\bar{n}^\mu}{\overrightarrow{\partial_-}}
        -\frac{n^\mu}{\overleftarrow{\partial_+}}\biggr)
    \mathbb{O}_{21,n}^{li}
    \biggr]
    \nonumber
    \\
    &\quad
    +\frac{\img g}{N_c}\,
    (\gamma_T^\mu)_{ij} \delta_{kl}
    \biggl[
    \mathbb{O}_{12,\bar{n}}^{li}
    \biggl(\frac{\bar{n}^\nu}{\overrightarrow{\partial_-}}
        -\frac{n^\nu}{\overleftarrow{\partial_+}}\biggr)
    \overline{\mathcal{O}}_{11,n}^{jk}
    -
    \overline{\mathbb{O}}_{12,\bar{n}}^{jk}
    \biggl(\frac{\bar{n}^\nu}{\overrightarrow{\partial_-}}
        -\frac{n^\nu}{\overleftarrow{\partial_+}}\biggr)
    \mathcal{O}_{11,n}^{li}
    \nonumber
    \\
    &\quad\qquad\quad
    +
    \mathcal{O}_{11,\bar{n}}^{li}
    \biggl(\frac{\bar{n}^\nu}{\overrightarrow{\partial_-}}
        -\frac{n^\nu}{\overleftarrow{\partial_+}}\biggr)
    \overline{\mathbb{O}}_{12,n}^{jk}
    -
    \overline{\mathcal{O}}_{11,\bar{n}}^{jk}
    \biggl(\frac{\bar{n}^\nu}{\overrightarrow{\partial_-}}
        -\frac{n^\nu}{\overleftarrow{\partial_+}}\biggr)
    \mathbb{O}_{12,n}^{li}
    \biggr]\,,
    \label{eq:WNLP}
\end{align}
where again we have considered the leading perturbative order and have suppressed the space-time arguments of the operators for brevity.  The non-trivial Wilson coefficients that enter beyond leading order in perturbation theory involve a convolution in the light-cone positions, so we prefer to include them only when switching to momentum space in \eq{W_intermediate}. As explained in more detail in ref.~\cite{Vladimirov:2021hdn}, two different types of power corrections can be distinguished in \eq{WNLP}, those involving derivatives of twist-2 operators (kinematic power corrections) and genuine twist-3 contributions.

At this point we are ready to write the hadronic tensor in momentum space, including the Wilson coefficients and external states, allowing us to pass from operators to jet functions.  To switch to momentum-space, we note that the inverse derivatives that appear in eq.~(\ref{eq:WNLP}) can be treated using the following identity,
\begin{align} \label{eq:inv_deriv}
    \int\frac{\df y^\pm}{2\pi}\,e^{\img  y^\pm q^\mp}\,
    \frac{1}{\partial_\mp} f(y^\pm)
    &=
    \frac{\img}{q^\mp}\,
    \int\frac{\df y^\pm}{2\pi}\,
    e^{\img  y^\pm q^\mp}\,
    f(y^\pm)\,.
\end{align}
Since the background fields of the different collinear sectors are not coupled at this order in the power counting, the matrix elements factorize and the result for the hadronic tensor can be written in terms of products of jet functions, 
\begin{align} \label{eq:W_intermediate}
    W_{\text{LP}}^{\mu\nu}(q)
    &=
    \frac{1}{N_c} (\gamma_T^\mu)_{ij} (\gamma_T^\nu)_{kl}\,
    \bigl|C_1(2q^+ q^-)\bigr|^2
    \nn \\
    &\qquad\times
    \int\frac{\df^2 b}{(2\pi)^2}\,e^{\img b\cdot q}
    \Bigl\{
    \bigl[\mathscr{J}^{\bar{q}}_{11,\bar{n}}(b)\bigr]^{li}
    \bigl[\mathscr{J}^q_{11,n}(b)\bigr]^{jk}
    +
    \bigl[\mathscr{J}^q_{11,\bar{n}}(b)\bigr]^{jk}
    \bigl[\mathscr{J}^{\bar{q}}_{11,n}(b)\bigr]^{li}
    \Bigr\}\,.    
    \\[2ex]
    W_{\text{NLP}}^{\mu\nu}(q)
    &=
    -\frac{\img}{N_c} \frac{1}{q^+}
    \bigl[n^\mu (\gamma_T^\rho)_{ij} (\gamma_T^\nu)_{kl} 
        + n^\nu (\gamma_T^\mu)_{ij} (\gamma_T^\rho)_{kl}\bigr]\,
    \bigl|C_1(2q^+ q^-)\bigr|^2
    \nn \\
    &\quad\qquad\times
    \int\frac{\df^2 b}{(2\pi)^2}\,e^{\img  b\cdot q}
    \Bigl\{
    \partial_\rho \bigl[\mathscr{J}^{\bar{q}}_{11,\bar{n}}(b)\bigr]^{li}\,
    \bigl[\mathscr{J}^q_{11,n}(b)\bigr]^{jk}
    +
    \partial_\rho \bigl[\mathscr{J}^q_{11,\bar{n}}(b)\bigr]^{jk}\,
    \bigl[\mathscr{J}^{\bar{q}}_{11,n}(b)\bigr]^{li}
    \Bigr\}
    \nonumber
    \\
    &\quad
    -
    \frac{\img}{N_c}
    \frac{1}{q^-}
    \bigl[\bar{n}^\mu (\gamma_T^\rho)_{ij} (\gamma_T^\nu)_{kl} 
        + \bar{n}^\nu (\gamma_T^\mu)_{ij} (\gamma_T^\rho)_{kl}\bigr]\,
    \bigl|C_1(2q^+ q^-)\bigr|^2
    \nonumber
    \\
    &\quad\qquad\times
    \int\frac{\df^2 b}{(2\pi)^2}\,e^{\img  b\cdot q}
    \Bigl\{
    \bigl[\mathscr{J}^{\bar{q}}_{11,\bar{n}}(b)\bigr]^{li}\,
    \partial_\rho \bigl[\mathscr{J}^q_{11,n}(b)\bigr]^{jk}
    +
    \bigl[\mathscr{J}^q_{11,\bar{n}}(b)\bigr]^{jk}\,
    \partial_\rho \bigl[\mathscr{J}^{\bar{q}}_{11,n}(b)\bigr]^{li}
    \Bigr\}
    \nonumber
    \\
    &\quad
    +\frac{\img}{N_c}\,
    \biggl[\frac{\bar{n}^\mu}{q^-} - \frac{n^\mu}{q^+}\biggr]
    \delta_{ij} (\gamma_T^\nu)_{kl}
    \int_0^1\df \xi\, C_2^*(\xi;2q^+ q^-) C_1(2q^+ q^-)
    \nonumber
    \\
    &\quad\qquad\times
    \int\frac{\df^2 b}{(2\pi)^2}\,e^{\img  b\cdot q}
    \Bigl\{
    \bigl[\mathscr{J}^{\bar{q}}_{21,\bar{n}}(\xi,b)\bigr]^{li}\,
    \bigl[\mathscr{J}^q_{11,n}(b)\bigr]^{jk}
    -
    \bigl[\mathscr{J}^q_{21,\bar{n}}(\xi,b)\bigr]^{jk}\,
    \bigl[\mathscr{J}^{\bar{q}}_{11,n}(b)\bigr]^{li}
    \nonumber
    \\
    &\quad\qquad\quad
    +
    \bigl[\mathscr{J}^{\bar{q}}_{11,\bar{n}}(b)\bigr]^{li}\,
    \bigl[\mathscr{J}^q_{21,n}(\xi,b)\bigr]^{jk}
    -
    \bigl[\mathscr{J}^q_{11,\bar{n}}(b)\bigr]^{jk}\,
    \bigl[\mathscr{J}^{\bar{q}}_{21,n}(\xi,b)\bigr]^{li}
    \Bigr\}
    \nonumber
    \\
    &\quad
    +\frac{\img}{N_c}\,
    \biggl[\frac{\bar{n}^\nu}{q^-} - \frac{n^\nu}{q^+}\biggr]
    (\gamma_T^\mu)_{ij} \delta_{kl}
    \int_0^1\df \xi\, C_1^*(q^+ q^-) C_2(\xi;2q^+ q^-)
    \nonumber
    \\
    &\quad\qquad\times
    \int\frac{\df^2 b}{(2\pi)^2}\,e^{\img  b\cdot q}
    \Bigl\{
    \bigl[\mathscr{J}^{\bar{q}}_{12,\bar{n}}(\xi,b)\bigr]^{li}\,
    \bigl[\mathscr{J}^q_{11,n}(b)\bigr]^{jk}
    -
    \bigl[\mathscr{J}^q_{12,\bar{n}}(\xi,b)\bigr]^{jk}\,
    \bigl[\mathscr{J}^{\bar{q}}_{11,n}(b)\bigr]^{li}
    \nonumber
    \\
    &\quad\qquad\quad
    +\bigl[\mathscr{J}^{\bar{q}}_{11,\bar{n}}(b)\bigr]^{li}\,
    \bigl[\mathscr{J}^q_{12,n}(\xi,b)\bigr]^{jk}
    -
    \bigl[\mathscr{J}^q_{11,\bar{n}}(b)\bigr]^{jk}\,
    \bigl[\mathscr{J}^{\bar{q}}_{12,n}(\xi,b)\bigr]^{li}
    \Bigr\}\,.
\end{align}
Here, $C_1$ and $C_2$ are matching coefficients that can be calculated in perturbation theory. These coefficients arise when the current operator is written in terms of the background fields, and we follow the definitions of~\cite{Vladimirov:2021hdn},  and the symbol $\mathscr{J}$  denotes jet functions with open spinor indices. At leading order in perturbation theory $C_i=1$, consistent with eqs.~(\ref{eq:WLP}) and (\ref{eq:WNLP}).
 The coefficient $C_1$ is known up to three-loops~\cite{Gehrmann:2010ue} (see also~\cite{Das:2019btv,Moch:2021qrk}), while the coefficient $C_2$ is calculated up to NLO in~\cite{Vladimirov:2021hdn}, section 6. For definiteness, the $C_2$ in eq.~(\ref{eq:W_intermediate}) is related to the one in ref.~\cite{Vladimirov:2021hdn}
by $C_2(\xi, 2 q^+q^-)=C_2^{\textbf{\cite{Vladimirov:2021hdn}}}(x_1=\xi,x_2=-\xi)$.
The bounds of the $\xi$ integral arise from the range over which the jet functions have support.

The jet functions that appear above are defined in terms of matrix elements of the building-block operators in \eqs{U1}{U2}. The twist-2 jet functions are given by
\begin{align} \label{eq:J_11_indices}
    \bigl[\mathscr{J}^q_{11,\bar{n}}(b)\bigr]^{jk}
    &=
    \tr_c \int\frac{\df y^-}{2\pi}\,e^{\img y^- q^+}\,
    \bra{0}
    \bigl[U_{1,\bar{n}}(y^-,b)\bigr]_j^{(-)}
    \ket{J^{\bar{n}}_\text{alg}}
    \bra{J^{\bar{n}}_\text{alg}}
    \bigl[\overline{U}_{1,\bar{n}}(0,0)\bigr]_k^{(+)}
    \ket{0}\,,
    \nn \\ \nn
    \bigl[\mathscr{J}^q_{11,n}(b)\bigr]^{jk}
    &=
    \tr_c \int\frac{\df y^+}{2\pi}\,e^{\img y^+ q^-}\,
    \bra{0}
    \bigl[U_{1,n}(y^+,b)\bigr]_j^{(-)}
    \ket{J^{n}_\text{alg}}
    \bra{J^{n}_\text{alg}}
    \bigl[\overline{U}_{1,n}(0,0)\bigr]_k^{(+)}
    \ket{0}\,,
    \\[2ex]\nn
    \bigl[\mathscr{J}^{\bar{q}}_{11,\bar{n}}(b)\bigr]^{li}
    &=
    \int\frac{\df y^-}{2\pi}\,e^{\img y^- q^+}\,
    \bra{0}
    \bigl[\overline{U}_{1,\bar{n}}(y^-,b)\bigr]_i^{(-)}
    \ket{J^{\bar{n}}_\text{alg}}
    \bra{J^{\bar{n}}_\text{alg}}
    \bigl[U_{1,\bar{n}}(0,0)\bigr]_l^{(+)}
    \ket{0}\,,
    \\ 
    \bigl[\mathscr{J}^{\bar{q}}_{11,n}(b)\bigr]^{li}
    &=
    \int\frac{\df y^+}{2\pi}\,e^{\img y^+ q^-}\,
    \bra{0}
    \bigl[\overline{U}_{1,n}(y^+,b)\bigr]_i^{(-)}
    \ket{J^{n}_\text{alg}}
    \bra{J^{n}_\text{alg}}
    \bigl[U_{1,n}(0,0)\bigr]_l^{(+)}
    \ket{0}\,,
\end{align}
and the twist-3 jet functions are defined as 
\begin{align}  \label{eq:J_21_indices}
    \bigl[\mathscr{J}^q_{21,\bar{n}}(\xi,b)\bigr]^{jk}
    &=
    \frac{1}{\xi - \img\delta_E^+}
    \,\tr_c \int\frac{\df y_1^-}{2\pi}\frac{\df y_2^-}{2\pi}\,
    e^{ \img  (\bar{\xi} y_1^- + \xi y_2^-) q^+}\,
    \nn \\ \nn
    &\qquad\times
    \bra{0}
    \bigl[U_{2,\bar{n}}(\{y_1^-,y_2^-\},b)\bigr]^{(-)}_j
    \ket{J^{\bar{n}}_\text{alg}}
    \bra{J^{\bar{n}}_\text{alg}}
    \bigl[\overline{U}_{1,\bar{n}}(0,0)\bigr]_k^{(+)}
    \ket{0}\,,
    \nonumber
    \\ \nn
    \bigl[\mathscr{J}^q_{21,n}(\xi,b)\bigr]^{jk}
    &=
    \frac{1}{\xi - \img\delta_E^-}
    \,\tr_c\int\frac{\df y_1^+}{2\pi}\frac{\df y_2^+}{2\pi}\,
    e^{ \img  (\bar{\xi} y_1^+ + \xi y_2^+) q^-}\,
    \\ \nn
    &\qquad\times
    \bra{0}
    \bigl[U_{2,n}(\{y_1^+,y_2^+\},b)\bigr]^{(-)}_j
    \ket{J^n_\text{alg}}
    \bra{J^n_\text{alg}}
    \bigl[\overline{U}_{1,n}(0,0)\bigr]_k^{(+)}
    \ket{0}\,,
    \nonumber
    \\[2ex]\nn
    \bigl[\mathscr{J}^{\bar{q}}_{21,\bar{n}}(\xi,b)\bigr]^{li}
    &=
    \frac{1}{\xi - \img\delta_E^+}
    \int\frac{\df y_1^-}{2\pi}\frac{\df y_2^-}{2\pi}\,
    e^{ \img  (\bar{\xi} y_1^- + \xi y_2^-) q^+}\,
    \\ \nn
    &\qquad\times
    \bra{0}
    \bigl[\overline{U}_{2,\bar{n}}(\{y_1^-,y_2^-\},b)\bigr]_i^{(-)}
    \ket{J^{\bar{n}}_\text{alg}}
    \bra{J^{\bar{n}}_\text{alg}}
    \bigl[U_{1,\bar{n}}(0,0)\bigr]_l^{(+)}
    \ket{0}\,,
    \nonumber
    \\ \nn
    \bigl[\mathscr{J}^{\bar{q}}_{21,n}(\xi,b)\bigr]^{li}
    &=
    \frac{1}{\xi - \img\delta_E^-}
    \int\frac{\df y_1^+}{2\pi}\frac{\df y_2^+}{2\pi}\,
    e^{ \img  (\bar{\xi} y_1^+ + \xi y_2^+) q^-}\,
    \\ \nn
    &\qquad\times
    \bra{0}
    \bigl[\overline{U}_{2,n}(\{y_1^+,y_2^+\},b)\bigr]_i^{(-)}
    \ket{J^n_\text{alg}}
    \bra{J^n_\text{alg}}
    \bigl[U_{1,n}(0,0)\bigr]_l^{(+)}
    \ket{0}\,,
    \nonumber
    \\[2ex] \nn
    \bigl[\mathscr{J}^q_{12,\bar{n}}(\xi,b)\bigr]^{jk}
    &=
    -\frac{1}{\xi + \img\delta_E^+}
    \, \tr_c \int\frac{\df y^-}{2\pi}\frac{\df z^-}{2\pi}\,
    e^{ \img  (y^- - \xi z^-) q^+}\,
    \\ \nn
    &\qquad\times
    \bra{0}
    \bigl[U_{1,\bar{n}}(y^-,b)\bigr]^{(-)}_j
    \ket{J^{\bar{n}}_\text{alg}}
    \bra{J^{\bar{n}}_\text{alg}}
    \bigl[\overline{U}_{2,\bar{n}}(\{0,z^-\},0)\bigr]_k^{(+)}
    \ket{0}\,,
    \nonumber
    \\ \nn
    \bigl[\mathscr{J}^q_{12,n}(\xi,b)\bigr]^{jk}
    &=
    -\frac{1}{\xi + \img\delta_E^-}
    \, \tr_c \int\frac{\df y^+}{2\pi}\frac{\df z^+}{2\pi}\,
    e^{ \img  (y^+ - \xi z^+) q^-}\,
    \\ \nn
    &\qquad\times
    \bra{0}
    \bigl[U_{1,n}(y^+,b)\bigr]^{(-)}_j
    \ket{J^n_\text{alg}}
    \bra{J^n_\text{alg}}
    \bigl[\overline{U}_{2,n}(\{0,z^+\},0)\bigr]_k^{(+)}
    \ket{0}\,,
    \nonumber
    \\[2ex]\nn
    \bigl[\mathscr{J}^{\bar{q}}_{12,\bar{n}}(\xi,b)\bigr]^{li}
    &=
    -\frac{1}{\xi + \img\delta_E^+}
    \int\frac{\df y^-}{2\pi}\frac{\df z^-}{2\pi}\,
    e^{ \img  (y^- - \xi z^-) q^+}\,
    \\ \nn
    &\qquad\times
    \bra{0}
    \bigl[\overline{U}_{1,\bar{n}}(y^-,b)\bigr]_i^{(-)}
    \ket{J^{\bar{n}}_\text{alg}}
    \bra{J^{\bar{n}}_\text{alg}}
    \bigl[U_{2,\bar{n}}(\{0,z^-\},0)\bigr]_l^{(+)}
    \ket{0}\,,
    \nonumber
    \\ \nn
    \bigl[\mathscr{J}^{\bar{q}}_{12,n}(\xi,b)\bigr]^{li}
    &=
    -\frac{1}{\xi + \img\delta_E^-}
    \int\frac{\df y^+}{2\pi}\frac{\df z^+}{2\pi}\,
    e^{ \img  (y^+ - \xi z^+) q^-}\,
    \\ 
    &\qquad\times
    \bra{0}
    \bigl[\overline{U}_{1,n}(y^-,b)\bigr]_i^{(-)}
    \ket{J^n_\text{alg}}
    \bra{J^n_\text{alg}}
    \bigl[U_{2,n}(\{0,z^+\},0)\bigr]_l^{(+)}
    \ket{0}\,,    
\end{align}
and we have introduced for convenience
\begin{align}
    \bar\xi&=1-\xi\,.
\end{align}
Note that whether the operator appears in the amplitude or conjugate amplitude is determined by whether the operator is causal or anti-causal. From now on we will drop these superscripts on the operators. 

\subsection{Jet function definitions}
\label{sec:JF}

The next step in the factorization is to define a minimal set of (bare) jet functions that have no open spinor or Lorentz indices,
as the set of jet functions in sec.~\ref{sec:EHT} contains redundancy.

First, to get rid off the open spinor indices, we apply the Fierz relations in app.~\ref{sec:Fierz}. This gives rise to four twist-2 jet functions with no Lorentz index
\begin{align}
    (\gamma^+)_{kj}
    \bigl[\mathscr{J}^q_{11,\bar{n}}(b)\bigr]^{jk}
    &=
    2N_c\, J^q_{11,\bar{n}}(b)\,,
    &
    (\gamma^+)_{il}
    \bigl[\mathscr{J}^{\bar{q}}_{11,\bar{n}}(b)\bigr]^{li}
    &=
    2N_c\, J^{\bar{q}}_{11,\bar{n}}(b)\,,
    \nn \\
    (\gamma^-)_{kj}
    \bigl[\mathscr{J}^q_{11,n}(b)\bigr]^{jk}
    &=
    2N_c\, J^q_{11,n}(b)\,,
    &
    (\gamma^-)_{il}
    \bigl[\mathscr{J}^{\bar{q}}_{11,n}(b)\bigr]^{li}
    &=
    2N_c\, J^{\bar{q}}_{11,n}(b)\,,
\end{align}
and eight twist-3 jet functions with one open Lorentz index, 
\begin{align}
    +\epsilon_T^{\mu\alpha}(\sigma^{\alpha+} \gamma^5)_{kj}
    \bigl[\mathscr{J}^q_{21,\bar{n}}(\xi,b)\bigr]^{jk}
    &=
    2N_c\, J^{\mu,q}_{21,\bar{n}}(\xi,b)\,,
    &
    +\epsilon_T^{\mu\alpha}(\sigma^{\alpha+} \gamma^5)_{il}
    \bigl[\mathscr{J}^{\bar{q}}_{21,\bar{n}}(\xi,b)\bigr]^{li}
    &=
    2N_c\, J^{\mu,\bar{q}}_{21,\bar{n}}(\xi,b)\,,
    \nonumber
    \\[1ex]
    -\epsilon_T^{\mu\alpha}(\sigma^{\alpha-} \gamma^5)_{kj}
    \bigl[\mathscr{J}^q_{21,n}(\xi,b)\bigr]^{jk}
    &=
    2N_c\, J^{\mu,q}_{21,n}(\xi,b)\,,
    &
    -\epsilon_T^{\mu\alpha}(\sigma^{\alpha-} \gamma^5)_{il}
    \bigl[\mathscr{J}^{\bar{q}}_{21,n}(\xi,b)\bigr]^{li}
    &=
    2N_c\, J^{\mu,\bar{q}}_{21,n}(\xi,b)\,,
    \nonumber
    \\[3ex]
    -\epsilon_T^{\mu\alpha}(\sigma^{\alpha+} \gamma^5)_{kj}
    \bigl[\mathscr{J}^q_{12,\bar{n}}(\xi,b)\bigr]^{jk}
    &=
    2N_c\, J^{\mu,q}_{12,\bar{n}}(\xi,b)\,,
    &
    -\epsilon_T^{\mu\alpha}(\sigma^{\alpha+} \gamma^5)_{il}
    \bigl[\mathscr{J}^{\bar{q}}_{12,\bar{n}}(\xi,b)\bigr]^{li}
    &=
    2N_c\, J^{\mu,\bar{q}}_{12,\bar{n}}(\xi,b)\,,
    \nonumber
    \\[1ex]
    +\epsilon_T^{\mu\alpha}(\sigma^{\alpha-} \gamma^5)_{kj}
    \bigl[\mathscr{J}^q_{12,n}(\xi,b)\bigr]^{jk}
    &=
    2N_c\, J^{\mu,q}_{12,n}(\xi,b)\,,
    &
    +\epsilon_T^{\mu\alpha}(\sigma^{\alpha-} \gamma^5)_{il}
    \bigl[\mathscr{J}^{\bar{q}}_{12,n}(\xi,b)\bigr]^{li}
    &=
    2N_c\, J^{\mu,\bar{q}}_{12,n}(\xi,b)\,.
    \nn\\
    \label{eq:J8}
\end{align}
The factor $2N_c$ accounts for averaging over the spin and color of the initial quark field. The different overall sign between $J_{21,n}$ and $J_{21,\bar n}$ is chosen because $\eps_T^{\mu\nu} = \eps^{\mu\nu\rho\si} n_\rho \bar n_\si$ changes sign when  $n \leftrightarrow \bar n$.

Discrete symmetries provide an additional  reduction of independent jet functions. Jet algorithms are spin-independent and respect the $C$,$P$ and $T$ symmetries of QCD. This allows us to derive that the LP jet functions satisfy,
\begin{align}
    J^{\bar{q}}_{11,\bar{n}}(b)
    &=
    J^{q}_{11,\bar{n}}(b)\,,
    &
    J^{\bar{q}}_{11,n}(b)
    &=
    J^{q}_{11,n}(b)\,,
\end{align}
reducing the number of independent twist-2 jet functions down to two, one for $n$ and one for the $\bar n$ direction. For the twist-3 jet functions we have
\begin{align} \label{eq:J_rel}
    J^{\rho,q}_{12,\bar{n}}(\xi,b)
    &=
    -\bigl[J^{\rho,q}_{21,\bar{n}}(\xi,-b)\bigr]^*\,,
    &
    J^{\rho,\bar{q}}_{21,\bar{n}}(\xi,b)
    &=
    J^{\rho,q}_{21,\bar{n}}(\xi,b)\,,
    \nn \\
    J^{\rho,q}_{12,n}(\xi,b)
    &=
    -\bigl[J^{\rho,q}_{21,n}(\xi,-b)\bigr]^*\,,
    &
    J^{\rho,\bar{q}}_{21,n}(\xi,b)
    &=
    J^{\rho,q}_{21,n}(\xi,b)\,,
\end{align}
and similarly for $J_{12}^{\bar q}$. This reduces the number of independent twist-3 jet functions from eight down to two. Since the quark and anti-quark distributions are identical we will drop the corresponding label from now on. The minus signs appearing in the complex conjugation cancel when performing the decomposition in~\eq{J_ld}. Thus the signs in eq.~(\ref{eq:J8}) were chosen precisely such that these symmetry relations do not contain minus signs.

Both the NLP jet functions and the derivatives of the LP jet functions that appear in the hadronic tensor carry an open (transverse) Lorentz index. To achieve a fully factorized formula for the hadronic tensor we must decompose these objects into the available structures $b^\mu$ and $\epsilon_T^{\mu\nu}b_\nu$. Parity implies that only $b^\mu$ contributes, leading us to define
\begin{align} \label{eq:J_ld}
    J^{\rho}_{21,\bar{n}}(\xi,b)
    &=\frac{b^\rho}{b^2} J_{21,\bar{n}}(\xi,\mathbf{b}^2)\,,
    &
    \partial_\rho J_{11,\bar{n}}(b)
    &=\frac{b_\rho}{b^2} J^{\prime}_{11,\bar{n}}(\mathbf{b}^2)\,,
    \nn \\ 
    J^{\rho}_{21,n}(\xi,b)
    &=\frac{b^\rho}{b^2} J_{21,n}(\xi,\mathbf{b}^2)\,,
    &
    \partial_\rho J^q_{11,n}(b)
    &=\frac{b_\rho}{b^2} J^{\prime\,q}_{11,n}(\mathbf{b}^2)\,.
\end{align}
This final step removes all open Lorentz indices from the jet definitions. Since all functions now only depend on the magnitude of $b$, we can perform the angular integration using the following identities,
\begin{align}
    \int\frac{\df^2 b}{(2\pi)^2}\,e^{\img b\cdot q}\,f(\mathbf{b}^2)
    &=
    \frac{1}{2}\int_0^\infty\df|\mathbf{b}|^2\,
    \frac{J_0(|\mathbf{b}||\mathbf{q}|)}{2\pi}\,
    f(\mathbf{b}^2)\,,
    \nn \\ 
    \int\frac{\df^2 b}{(2\pi)^2}\,e^{\img b\cdot q}\,
    \frac{\img b^\rho}{b^2}\,f(\mathbf{b}^2)
    &=
    \frac{q_T^\rho}{2}\int_0^\infty\df|\mathbf{b}|^2\,
    \frac{J_1(|\mathbf{b}||\mathbf{q}|)}{2\pi|\mathbf{b}||\mathbf{q}|}\,
    f(\mathbf{b}^2)\,.
\end{align}

Applying the above decomposition to the definitions presented in \eqs{J_11_indices}{J_21_indices}, the final expressions for the jet functions read,
\begin{align}\label{eq:Jbare_final}
    J^\text{bare}_{11,\bar{n}}(\mathbf{b}^2)
    &=
    \frac{1}{2N_c}\,\int\frac{\df y^-}{2\pi}\,e^{\img y^- q^+}\,
    \tr\bigl[\gamma^+\bra{0}U_{1,\bar{n}}(y^-,b)
    \ket{J^{\bar{n}}_\text{alg}}
    \bra{J^{\bar{n}}_\text{alg}}
    \overline{U}_{1,\bar{n}}(0,0)
    \ket{0}\bigr]\,,
    \\
    J^{\prime\,\text{bare}}_{11,\bar{n}}(\mathbf{b}^2)
    &=
    b^\mu\frac{\partial}{\partial b^\mu} J_{11,\bar{n}}^{\rm bare}(\mathbf{b}^2)\,,
    \\
    J^\text{bare}_{21,\bar{n}}(\xi,\mathbf{b}^2)
    &=
    n^\mu \bar{n}^\nu b^\rho\epsilon_{\mu\nu\rho\sigma}\,
    \frac{1}{\xi - \img \delta_E^+}\,
    \frac{1}{2N_c}
    \int\frac{\df y_1^-}{2\pi}\frac{\df y_2^-}{2\pi}\,
    e^{ \img  (\bar{\xi} y_1^- + \xi y_2^-) q^+}\,
    \nn\\
    &\qquad\times
    \tr\Bigl[\sigma^{\sigma+} \gamma^5
    \bra{0}
    U_{2,\bar{n}}(\{y_1^-,y_2^-\},b)
    \ket{J^{\bar{n}}_\text{alg}}
    \bra{J^{\bar{n}}_\text{alg}}
    \overline{U}_{1,\bar{n}}(0,0)
    \ket{0}\Bigr]\,,
\end{align}
where the trace is over both the color and Dirac indices. The jet functions for the $n$-collinear sector can be obtained by replacing $n\leftrightarrow\bar{n}$.

\subsection{Factorization}
\label{sec:factorization}

Starting from eq.~\eqref{eq:W_intermediate}, we now apply the relations and definitions of the previous section to obtain a factorized expression for the hadronic tensor,

\begin{align} \label{eq:W_final}
    W_{\text{LP}}^{\mu\nu}(q)
    &=
    N_c\, (-g_T^{\mu\nu})
    H_1(Q^2)
    \int\df|\mathbf{b}|^2\,
    \frac{J_0(|\mathbf{b}||\mathbf{q}|)}{2\pi}
    J_{11,\bar{n}}(\mathbf{b}^2)\,J_{11,n}(\mathbf{b}^2)\,,    
    \\[2ex]
    W_{\text{NLP}}^{\mu\nu}(q) 
    &=
    N_c\,\biggl[\frac{n^\mu q_T^\nu}{q^+}
        +\frac{n^\nu q_T^\mu}{q^+}\biggr]\,
    H_1(Q^2)
    \int_0^\infty\df|\mathbf{b}|^2\,
    \frac{J_1(|\mathbf{b}||\mathbf{q}|)}{2\pi|\mathbf{b}||\mathbf{q}|}\,
    J^\prime_{11,\bar{n}}(\mathbf{b}^2)\,J_{11,n}(\mathbf{b}^2)
    \nonumber
    \\
    &\quad
    +N_c\,\biggl[\frac{\bar{n}^\mu q_T^\nu}{q^-}
        +\frac{\bar{n}^\nu q_T^\mu}{q^-}\biggr]\,
    H_1(Q^2)
    \int_0^\infty\df|\mathbf{b}|^2\,
    \frac{J_1(|\mathbf{b}||\mathbf{q}|)}{2\pi|\mathbf{b}||\mathbf{q}|}\,
    J_{11,\bar{n}}(\mathbf{b}^2)\,J^\prime_{11,n}(\mathbf{b}^2)
    \nonumber
    \\
    &\quad
    +N_c\,\,
    \biggl[\frac{\bar{n}^\mu q_T^\nu}{q^-} - \frac{n^\mu q_T^\nu}{q^+}\biggr]
    \int_0^1\df\xi\, H_2(\xi,Q^2)
    \nonumber
    \\
    &\quad\qquad\times
    \int_0^\infty\df|\mathbf{b}|^2\,
    \frac{J_1(|\mathbf{b}||\mathbf{q}|)}{2\pi|\mathbf{b}||\mathbf{q}|}\,
    \Bigl\{
    J_{21,\bar{n}}(\xi,\mathbf{b}^2) J_{11,n}(\mathbf{b}^2)
    -J_{11,\bar{n}}(\mathbf{b}^2) J_{21,n}(\xi,\mathbf{b}^2)
    \Bigr\}
    \nonumber
    \\
    &\quad
    +N_c\,\,
    \biggl[\frac{\bar{n}^\nu q_T^\mu}{q^-} - \frac{n^\nu q_T^\mu}{q^+}\biggr]
    \int_0^1\df\xi\, H_2^*(\xi,Q^2)
    \nonumber
    \\
    &\quad\qquad\times
    \int_0^\infty\df|\mathbf{b}|^2\,
    \frac{J_1(|\mathbf{b}||\mathbf{q}|)}{2\pi|\mathbf{b}||\mathbf{q}|}\,
    \Bigl\{
    J^*_{21,\bar{n}}(\xi,\mathbf{b}^2) J_{11,n}(\mathbf{b}^2)
    -J_{11,\bar{n}}(\mathbf{b}^2) J^*_{21,n}(\xi,\mathbf{b}^2)
    \Bigr\}\,.
\nn \end{align}
Here, $H_1$ and $H_2$ are defined in terms of the Wilson coefficients and are referred to as the hard functions
\begin{align}
    H_1(Q^2)
    &=
    |C_1(Q^2)|^2
    \,, \nn \\
    H_2(\xi,Q^2)
    &=
    C_2^*(\xi,Q^2)\,C_1(Q^2)
\,.\end{align}
There are two things to note here. First, note that the hard function $H_2$ is complex valued for the process we consider here. Second, we have replaced the arguments $2q^+ q^-$ of the Wilson coefficients by $Q^2$, as this replacement can be made up to power corrections (these are kinematic power corrections, see ref.~\cite{Vladimirov:2023aot}). For this latter point, it has been shown that higher-power corrections in fact lead to the replacement $2q^+ q^-\to Q^2$ in the Wilson coefficients, as expected by Lorentz invariance.

Next, we  combine the hadronic tensor with the leptonic tensor to obtain the cross section  for $e^+ e^-\to 2$ jets. In the above expression, the hadronic tensor is decomposed into several Lorentz structures. As an intermediate step, we record the following results for the contractions between these Lorentz structures and the leptonic tensor, 
\begin{alignat}{2}
    (-g_T^{\mu\nu})L_{\mu\nu}
    &=2Q^2\,(1-2y+2y^2)\,-\,4Q_T Q\,\cos(\phi_1)\,(1-2y)\sqrt{y(1-y)}
    +\mathcal{O}(Q_T^2)\,,
    \nn \\
    \frac{\bar{n}^\mu q_T^\nu}{q^-}L_{\mu\nu}
    &=
    -\frac{n^\mu q_T^\nu}{q^+}L_{\mu\nu}
    =4Q_T Q\,\cos(\phi_1)\,(1-2y)\sqrt{y(1-y)}
    +\mathcal{O}(Q_T^2)\,.
\end{alignat}
Note that the power-suppressed contributions coming from the two Lorentz structures are identical, making it experimentally impossible to distinguish between the different types of power corrections. Combining the full expression for the hadronic tensor of \eq{W_final} with the leptonic tensor in \eq{leptonic}, we find that the cross section is given by
\\[-2ex]

{\centering \fbox{\parbox{0.98\textwidth}{
\vspace{-2ex}
\begin{align}
\label{eq:sigmabare}
    \frac{\df\sigma}{\df y\, \df^2 {\bf q}} 
    &=
    \frac{\pi \alpha_{\rm em}^2Q_q^2 N_c}{Q^2}\,
    (2-4y+4y^2)
    \int_0^\infty\df|\mathbf{b}|^2\,
    \frac{J_0(|\mathbf{b}||\mathbf{q}|)}{2\pi}\,
    H_1(Q^2)\,J_{11,\bar{n}}(\mathbf{b}^2)\,J_{11,n}(\mathbf{b}^2)
    \nn \\
    &\quad
    -
    \frac{\pi \alpha_{\rm em}^2Q_q^2 N_c}{Q^2}\,
    \frac{Q_T}{Q}\,4\cos(\phi_1)\,(1-2y)\sqrt{y(1-y)}
    \nn\\
    &\quad\qquad\times
    \biggl\{
    \int_0^\infty\df|\mathbf{b}|^2\,
    \frac{J_0(|\mathbf{b}||\mathbf{q}|)}{2\pi}\,
    H_1(Q^2)\,J_{11,\bar{n}}(\mathbf{b}^2)\,J_{11,n}(\mathbf{b}^2)
    \nn\\
    &\quad\qquad\quad
    +2\int_0^\infty\df|\mathbf{b}|^2\,
    \frac{J_1(|\mathbf{b}||\mathbf{q}|)}{2\pi|\mathbf{b}||\mathbf{q}|}\,
    H_1(Q^2)\,
    \Bigl[J'_{11,\bar{n}}(\mathbf{b}^2)\,J_{11,n}(\mathbf{b}^2)
        -J_{11,\bar{n}}(\mathbf{b}^2)\,J'_{11,n}(\mathbf{b}^2)\Bigr]
    \nn\\
    &\quad\qquad\quad
    -2\int_0^\infty\df|\mathbf{b}|^2\,
    \frac{J_1(|\mathbf{b}||\mathbf{q}|)}{2\pi|\mathbf{b}||\mathbf{q}|}\,
    \int_0^1\df\xi\, H_2(\xi,Q^2)\,
    \nn\\
    &\quad\qquad\quad\qquad\times
    \Bigl[J_{21,\bar{n}}(\xi,\mathbf{b}^2)\,J_{11,n}(\mathbf{b}^2)
        -J_{11,\bar{n}}(\mathbf{b}^2)\,J_{21,n}(\xi,\mathbf{b}^2)\Bigr]
    +\text{c.c.}\,
    \biggr\}\,.
\end{align}
\vspace{-2ex}}}\\[2ex]} 
\noindent
Note that in the above expression the hard functions and jet functions are bare and unsubtracted quantities.

\section{Subtraction and renormalization}
\label{sec:Divergences}

In the previous section we arrived at a factorized expression for the cross section in terms of a set of jet functions and Wilson coefficients. We still need to treat the soft overlap between the collinear and anti-collinear region and renormalize all ingredients. Furthermore, the above result for the cross section contains some divergences, known in the literature as special rapidity divergences and endpoint divergences, that require treatment. In this section we discuss the overlap subtraction and the resulting cancellation of these divergences, as well as the renormalization of the jet functions. For other cases a scheme needs to be developed for treating endpoint divergences~\cite{Liu:2019oav,Liu:2020tzd,Beneke:2022obx}.

First, in sec.~\ref{sec:overlap}, we discuss the subtraction of the overlap region. In summary, the lack of a clear boundary between the collinear and anti-collinear regions gives rise to rapidity divergences, which cancel once the overlap region has been properly subtracted. In the method employed in this work, not all divergences cancel after the overlap region has been subtracted. These remaining special rapidity divergences already enter at lowest order in $a_s=\alpha_s/(4\pi)$, and are contained in the kinematic and higher-twist corrections. As a consequence, ill-defined convolutions between plus distributions and logarithms appear in the finite terms at higher orders in perturbation theory, which are  commonly referred to as endpoint divergences. We discuss the subtraction and cancellation of these divergences in sec.~\ref{sec:special}. The treatment of the overlap region is one of the main differences between the approach in SCET and the TMD operator expansion, which we compare in sec.~\ref{sec:discussion}. Finally, the renormalization of the jet functions and the corresponding evolution equations are discussed in sec.~\ref{sec:evolution}.

\subsection{Overlap subtraction}
\label{sec:overlap}

As mentioned in sec.~\ref{sec:OB}, the two collinear regions overlap in the soft region, which is doubly counted. The presence of this overlap has been extensively discussed in the literature about TMD factorization~\cite{Collins:2011zzd, Echevarria:2011epo,Chiu:2012ir}, and it has a similar treatment in the background field method~\cite{Vladimirov:2021hdn}. 

In the TMD operator expansion method, the overlap between the collinear and anti-collinear regions can be subtracted for bare operators by making the following replacements
\begin{align} \label{eq:overlap}
    \mathcal{O}_{11,\bar{n}}\overline{\mathcal{O}}_{11,n}\rightarrow
    \frac{\mathcal{O}_{11,\bar{n}}\overline{\mathcal{O}}_{11,n}}{\mathcal{S}_\text{LP}}\,,\quad
    \partial_\rho\mathcal{O}_{11,\bar{n}}
    \overline{\mathcal{O}}_{11,n}\rightarrow
    \frac{\partial_\rho\mathcal{O}_{11,\bar{n}}
    \overline{\mathcal{O}}_{11,n}}{\mathcal{S}_\text{LP}}\,,\quad
    \mathcal{O}_{21,\bar{n}}\overline{\mathcal{O}}_{11,n}\rightarrow
    \frac{\mathcal{O}_{21,\bar{n}}\overline{\mathcal{O}}_{11,n}}{\mathcal{S}_\text{LP}}\,,
\end{align}
with similar replacements for the other combinations of operators that appear in eq.~(\ref{eq:WNLP}). Using the $\delta$-regulator, the overlap factor is identical for both the leading twist and sub-leading twist operators, and is given by the soft function
\begin{align}
\label{eq:soft}
    \mathcal{S}_\text{LP}^{\rm bare} 
    (\mathbf{b}^2,\epsilon,2\delta^+\delta^-)
    &=
    \frac{1}{N_c}\tr_c\bigl\{
    \bra{0}[\infty \bar{n} + b,b][b,\infty n + b]
    [\infty n,0][0,\infty \bar{n}]\ket{0}
    \bigr\}\,.
\end{align}

The overlap subtraction as presented in \eq{overlap} happens at the level of the product of two operators from the different collinear sectors, and the rapidity divergences cancel between the $\mathcal{O}_n$, $\mathcal{O}_{\bar n}$ and $\mathcal{S}$. However, it is desirable to express the cross section in terms of ingredients that are each individually free of rapidity divergences. To define a rapidity-finite jet function, it is necessary to implement this subtraction by absorbing a square root of the soft function into each of the jet functions. To do this we use the fact that the soft function can be separated into two pieces,
\begin{align}
    \mathcal{S}_\text{LP}^{\rm bare}
    (\mathbf{b}^2,\epsilon,2\delta^+\delta^-)
    &=
    \sqrt{\mathcal{S}_\text{LP}^{\rm bare}
    \bigl(\mathbf{b}^2,\epsilon,(\delta_E^+)^2\zeta\bigr)}
    \sqrt{\mathcal{S}_\text{LP}^{\rm bare}
    \bigl(\mathbf{b}^2,\epsilon,(\delta_E^-)^2\zeta\bigr)}\,,
\end{align}
where 
\begin{align}
    \delta_E^+=\frac{\delta^+}{q^+}\,,\qquad
    \delta_E^-=\frac{\delta^-}{q^-}\,,\qquad\text{and}\qquad
    \zeta^2=(2q^+q^-)^2\,,
\end{align}
and $\zeta$ is referred to as the rapidity scale. Using the above identity, the overlap subtraction procedure can be implemented by making the following replacements
\begin{align}
\label{eq:Jbaresub}
    J^\text{bare}_{11,\bar{n}}(\mathbf{b}^2,\epsilon,\delta_E^+)
    &\rightarrow
    \frac{J^\text{bare}_{11,\bar{n}}(\mathbf{b}^2,\epsilon,\delta_E^+)}
    {\sqrt{\mathcal{S}_\text{LP}(\mathbf{b}^2,\epsilon,(\delta_E^+)^2\zeta)}}
    =
    J^\text{bare,sub}_{11,\bar{n}}(\mathbf{b}^2,\zeta,\epsilon)\,,
    \\
    J^{\prime\,\text{bare}}_{11,\bar{n}}(\mathbf{b}^2,\epsilon,\delta_E^+)
    &\rightarrow
    \frac{J^{\prime\,\text{bare}}_{11,\bar{n}}(\mathbf{b}^2,\epsilon,\delta_E^+)}
    {\sqrt{\mathcal{S}_\text{LP}(\mathbf{b}^2,\epsilon,(\delta_E^+)^2\zeta)}}\,,
    \\
    J^\text{bare}_{21,\bar{n}}(\xi,\mathbf{b}^2,\epsilon,\delta_E^+)
    &\rightarrow
    \frac{J^\text{bare}_{21,\bar{n}}(\xi,\mathbf{b}^2,\epsilon,\delta_E^+)}
    {\sqrt{\mathcal{S}_\text{LP}(\mathbf{b}^2,\epsilon,(\delta_E^+)^2\zeta)}}\,,
\end{align}
and similarly for the $n$-collinear direction.
After the subtraction procedure, the leading-power jet function $J^\text{bare,sub}_{11}$ is free of rapidity divergences but depends on the rapidity scale $\zeta$. The dependence on $\zeta$ will be discussed in \sec{evolution}.

\subsection{Special rapidity and endpoint divergences}
\label{sec:special}

The subtraction procedure of the TMD operator expansion method does not cancel all rapidity divergences. Specifically, $J^\prime_{11}$ and $J_{21}$, which respectively correspond to the the kinematical and higher-twist corrections, still contain divergences. These divergences can be subtracted and cancel in the cross section between the $n$ and the $\bar{n}$ sectors. We now discuss the cancellation of these divergences separately for the kinematic and higher-twist corrections.

\subsubsection*{Higher-twist correction}

For the twist-3 jet functions $J_{21}(\xi,\mathbf{b}^2)$ divergences appear in the convolution with $H_2$ in \eq{sigmabare} as $\xi\to0$. In most of the literature this is referred to as the problem of endpoint divergences, while in the TMD operator expansion literature these divergences are considered part of the special rapidity divergences. In the case of TMD factorization, different methods for treating these divergences exist and we compare the two in \sec{discussion}. In this work, we subtract the singular behavior of the jet functions as $\xi\to0$ by means of a sub-leading-power soft function. This ensures that the integral of the hard function and the twist-3 jet function is well defined. This was first discussed in refs.~\cite{Gao:2022ref,Gao:2023scet,Michel:2023esi}, which focused on SIDIS.

The key insight is that in the limit $\xi\to0$ the gluon field strength becomes soft. In the soft limit, the gluon decouples from the collinear sector and the field strength can be factored out of the collinear matrix element. As a result, the limit $\xi\to0$ is captured by
\begin{align} \label{eq:J_xi_to_0}
    \lim_{\xi\to0}
    J^\text{bare}_{21,\bar{n}}(\xi,\mathbf{b}^2)\,
    \mathcal{S}^\text{bare}_\text{LP}(\mathbf{b}^2)\,
    J^\text{bare}_{11,n}(\mathbf{b}^2)
    &=
    J^\text{bare}_{11,\bar{n}}(\mathbf{b}^2)
    S^\text{bare}_{21,\bar{n}}(\xi,\mathbf{b}^2)
    J^\text{bare}_{11,n}(\mathbf{b}^2)
    +\dotso\,,
\end{align}
where $\dotso$ denotes terms which are regular as $\xi\to0$ and a similar expression holds for the opposite collinear sector. 
In this expression the jet functions on the right-hand-side are the familiar twist-2 jet functions and $S_{21,\bar{n}}$ is an NLP soft function. The NLP soft functions for the two collinear sectors are defined as 
\begin{align}\label{eq:J_21_soft}
    S_{21,\bar{n}}^\text{bare}(\xi,\mathbf{b}^2,\epsilon,\delta_E^+,2\delta^+\delta^-)
    &=
    +\frac{b_\rho}{\xi-\img \delta_E^+}
    \int\frac{\df y_2^-}{2\pi}\,e^{i\xi y_2^- q^+}\,
    \frac{1}{N_c}\tr\Bigl[\bra{0}
    [b + y_2^- n + \infty n,b + y_2^- n]
    \\ & \quad \times
    [b + y_2^- n,b + y_2^- n + \infty\bar{n}]
    gF^{\rho+}(y_2^- n + b)\,
    \ket{X}\bra{X}
    [\infty \bar{n},0]
    [0,\infty n]
    \ket{0}\Bigr]\,,
    \nn \\[2ex]
    S_{21,n}^\text{bare}(\xi,\mathbf{b}^2,\epsilon,\delta_E^-,2\delta^+\delta^-)
    &=
    -\frac{b_\rho}{\xi-\img \delta_E^-}
    \int\frac{\df y_2^+}{2\pi}\,e^{i\xi y_2^+ q^-}\,
    \frac{1}{N_c}\tr\Bigl[\bra{0}
    gF^{\rho-}(y_2^+ \bar{n} + b)\,
    \nn\\
    &\quad\times
    [b + y_2^- n + \infty n,b + y_2^- n]
    [b + y_2^- n,b + y_2^- n + \infty\bar{n}]
    S_{\bar{n}}
    \ket{X}
    \nn\\
    &\quad\times
    \bra{X}
    [\infty \bar{n},0]
    [0,\infty n]
    \ket{0}\Bigr]\,.
\end{align}
This definition follows from replacing the collinear field strength operator in the $\mathbb{O}_{21}$ operators in \eq{WNLP} by a soft field strength, and following the same steps as in \sec{HadronicTensor}. 
Note that, strictly speaking, this derivation requires the introduction of a background field for the soft modes, although the soft function drops out in the final result for the NLP case we consider here.

Using \eq{J_xi_to_0}, we can subtract the singular behavior from the jet functions in the $\xi\to0$ limit
\begin{align}
    J^\text{bare,sub}_{21,\bar{n}}(\xi,\mathbf{b}^2,\zeta,\eps)
    &=
    \frac{J^\text{bare}_{21,\bar{n}}(\xi,\mathbf{b}^2)}
        {\sqrt{\mathcal{S}_\text{LP}
    (\mathbf{b}^2,\epsilon,(\delta_E^+)^2\zeta)}}
    -
    S^\text{bare}_{21,\bar{n}}(\xi,\mathbf{b}^2,\epsilon,\delta_E^+,\zeta)\,
    J^\text{bare,sub}_{11,\bar{n}}(\mathbf{b}^2,\zeta,\eps)\,.
\end{align}
Note that the subtracted jet function is regular as $\xi\to0$. As a result, its convolution with the hard function is well defined and free of any endpoint divergences.

The divergent $\ln \de_E^\pm$ and singular $1/\xi_+$ parts of the jet function are now contained in the subtraction terms, which cancel in the hadronic tensor between the two collinear sectors as the jet functions only appear in the combination
\begin{align*}
    J_{21,\bar{n}}(\xi,\mathbf{b}^2)\,J_{11,n}(\mathbf{b}^2)
    -J_{11,\bar{n}}(\mathbf{b}^2)\,J_{21,n}(\xi,\mathbf{b}^2)\,.
\end{align*}
For this cancellation to work, one needs the two soft functions in eq.~\eqref{eq:J_21_soft} to be identical, which is only the case if
\begin{align}
    \delta_E^+ = \delta_E^-\,,
\end{align}
see ref.~\cite{Ebert:2021jhy}.
Note that this constraint arises beyond leading order in $a_s$, where the hard function contains terms involving $\ln\xi$. Summarizing, the special rapidity and endpoint divergences in the higher-twist correction can be subtracted and cancel between the $n$ and $\bar{n}$ sector on the condition that the rapidity regulators for the two sectors are related.

\subsubsection*{Kinematical correction}

For the kinematical correction, only the following combination of jet functions appears after the subtraction procedure of sec.~\ref{sec:overlap} has been performed,
\begin{align}\label{eq:kin_combination}
    \frac{J^{\prime\,\text{bare}}_{11,\bar{n}}(\mathbf{b}^2,\epsilon,\delta_E^+)}
    {\sqrt{\mathcal{S}_\text{LP}
    (\mathbf{b}^2,\epsilon,(\delta_E^+)^2\zeta)}}\,
    J^\text{bare,sub}_{11,n}(\mathbf{b}^2,\zeta,\epsilon)
    -
    J^\text{bare,sub}_{11,\bar{n}}(\mathbf{b}^2,\zeta,\epsilon)\,
    \frac{J^{\prime\,\text{bare}}_{11,n}(\mathbf{b}^2,\epsilon,\delta_E^-)}
    {\sqrt{\mathcal{S}_\text{LP}
    (\mathbf{b}^2,\epsilon,(\delta_E^-)^2\zeta)}}\,.
\end{align}
Special rapidity divergences also appear in this expression because the derivative that defines $J^\prime_{11}$ is taken before the soft function is divided out. 

As for the higher-twist correction, the divergences in the above expression can be cancelled between the two collinear sectors. To see this we first define a subtracted jet function $J^\prime_{11}$, 
\begin{align}
    J^{\prime\,\text{bare,sub}}_{11,\bar{n}}
    (\mathbf{b}^2,\zeta,\epsilon)
    &=
    b^\rho\partial_\rho
    J^\text{bare,sub}_{11,\bar{n}}(\mathbf{b}^2,\zeta,\epsilon)\,.
\end{align}
that is free of any rapidity divergences, because we take the derivative after subtracting.
Using this definition, one can write
\begin{align}
    \frac{J^{\prime\,\text{bare}}_{11,\bar{n}}(\mathbf{b}^2,\epsilon,\delta_E^+)}
    {\sqrt{\mathcal{S}_\text{LP}
    (\mathbf{b}^2,\epsilon,(\delta_E^+)^2\zeta)}}
    &=
    J^{\prime\,\text{bare,sub}}_{11,\bar{n}}
    (\mathbf{b}^2,\zeta,\epsilon)
    +
    \frac{1}{2}
    J^\text{bare,sub}_{11,\bar{n}}(\mathbf{b}^2,\zeta,\epsilon)\,
    \frac{b^\rho\partial_\rho 
        \mathcal{S}_\text{LP}(\mathbf{b}^2,\epsilon,(\delta_E^+)^2\zeta)}
    {\mathcal{S}_\text{LP}(\mathbf{b}^2,\epsilon,(\delta_E^+)^2\zeta)}\,,
\end{align}
where the special rapidity divergences are contained in the second term. Inserting this in eq.~\eqref{eq:kin_combination}, we see that the rapidity divergent terms combine as follows,
\begin{align}
    \frac{b^\rho\partial_\rho 
        \mathcal{S}_\text{LP}(\mathbf{b}^2,\epsilon,(\delta_E^+)^2\zeta)}
    {\mathcal{S}_\text{LP}(\mathbf{b}^2,\epsilon,(\delta_E^+)^2\zeta)}
    -
    \frac{b^\rho\partial_\rho 
        \mathcal{S}_\text{LP}(\mathbf{b}^2,\epsilon,(\delta_E^-)^2\zeta)}
    {\mathcal{S}_\text{LP}(\mathbf{b}^2,\epsilon,(\delta_E^-)^2\zeta)}
    &=
    2b^\rho\partial_\rho\mathcal{D}(b)
    \ln\biggl(\frac{\delta_E^+}{\delta_E^-}\biggr)\,,
\end{align}
where the expression on the right-hand-side can be obtained using the exponentiation of the soft factor with $\mathcal{D}(b)$ the Collins-Soper kernel. We thus see that the special rapidity divergences for the kinematical corrections cancel. In fact if one takes $\delta_E^+=\delta_E^-$, as required by the cancellation of divergences for the higher-twist corrections, the subtraction terms cancel altogether.

\subsection{Discussion of special rapidity and endpoint divergences}
\label{sec:discussion}

The subtraction of the overlap region and the cancellation of special rapidity and endpoint divergences is treated differently in  SCET~\cite{Ebert:2021jhy,Gao:2022ref,Gao:2023scet,Michel:2023esi} and the TMD operator expansion literature~\cite{Vladimirov:2021hdn,Rodini:2022wic,Rodini:2022wki}. Here we compare the two approaches.

In SCET, a soft mode is included explicitly, and its contribution at next-to-leading power leads to a family of soft functions~\cite{Ebert:2021jhy}. At the level of bare matrix elements however, the contributions of these soft functions cancel in the cross section and as a result no NLP soft factor needs to be introduced. Nevertheless, these soft functions can be used to subtract the overlap regions and define subtracted TMD distributions at NLP~\cite{Gao:2022ref,Gao:2023scet,Michel:2023esi}. On the other hand, in the TMD operator expansion method, no soft mode is included and the overlap between the regions is subtracted by the same soft function that appears at leading power. The main consequence is that in this method, at NLP, not all rapidity divergences are cancelled by the subtraction procedure. The divergences that remain after the subtraction procedure are referred to as special rapidity divergences. This includes the singular behavior in the higher-twist corrections that appear as $\xi\to0$, which in the literature are commonly referred to as endpoint divergences.

For both approaches, the remaining divergences cancel in the cross section between the $n$ and $\bar{n}$ sectors. However, the subtraction and cancellation of the divergences is formulated differently between the two methods. In this work and in the SCET literature, it is formulated at all orders in perturbation theory by means of a soft matrix element. After the singular behavior of the jet functions has been subtracted, the divergent subtraction terms cancel in the combination
\begin{align}
    \int_0^1\df\xi\,
    H_2(\xi)\,
    J_{11,\bar{n}}(\mathbf{b}^2,\zeta)\,
    \frac{S_{21,\bar{n}}(\xi,\mathbf{b}^2,\delta_E^+,2\delta^+\delta^-) - S_{21,n}(\xi,\mathbf{b}^2,\delta_E^-,2\delta^+\delta^-)}{\mathcal{S}_\text{LP}(\mathbf{b}^2,2\delta^+\delta^-)}\,
    J_{11,n}(\mathbf{b}^2,\zeta)\,.
\end{align}
Since the hard function contains logarithms of $\xi$ at higher orders in perturbation theory, the only way for the divergences to cancel in the above expression is to have $\delta_E^+=\delta_E^-$. The above cancellation holds at the bare and renormalized level, however for the latter the distributions for the $n$ and $\bar{n}$ sectors have to be evolved together from the same initial to the same final scale. 

In refs.~\cite{Rodini:2022wic,Rodini:2022wki,Moos:2023yfa}, on the other hand, the cancellation of special rapidity divergences is formulated order-by-order in perturbation theory. In ref.~\cite{Rodini:2022wki} a subtracted distribution is defined by
\begin{align} \label{eq:J_21_alt}
    \boldsymbol{J}_{21}(\xi,\mathbf{b}^2)
    &=
    J_{21}(\xi,\mathbf{b}^2)
    -\bigl[\mathcal{R}_{21}\otimes J_{11}\bigr](\xi,\mathbf{b}^2)\,,
\end{align}
where $\mathcal{R}$ is to be constructed order-by-order in $\al_s$ such that the convolution between $\boldsymbol{J}_{21}$ and $H_2$ is finite. Note that we have converted their expression to our notation. At leading order in perturbation theory the hard function contains no terms involving $\ln\xi$, and so no constraint on the rapidity regulators arises. However, at higher orders in perturbation theory one would run into additional divergences coming from the integral of $\ln \xi$ in the hard function with the $1/(\xi - \img \de_E^\pm)$ in the jet functions, leading to $\ln^2 \de_E^\pm$. It is the cancellation of these additional divergences that constrains $\delta_E^+=\delta_E^-$. We therefore conjecture that if the approach of ref.~\cite{Rodini:2022wki} is extended to higher orders, one would find such a constraint.

\subsection{Renormalization and evolution} 
\label{sec:evolution}

The subtracted jet functions defined above are at the bare level and require renormalization. Since renormalization and evolution takes place at the operator level, it is independent of the recoil-free jet algorithm that is employed.

For the twist-2 jet functions the renormalization is multiplicative, and we define the renormalized jet function as
\begin{align}
    J_{11,\bar{n}}(\mathbf{b}^2,\mu,\zeta)
    &=
    Z_{J_{11}}(\zeta,\mu,\epsilon)\,
    J^\text{bare,sub}_{11,\bar{n}}
    (\mathbf{b}^2,\zeta,\epsilon)\,.
\end{align}
The renormalized jet functions depend on a renormalization scale $\mu$ and their evolution is given by
\begin{align}
    \frac{\df}{\df\ln\mu^2}J_{11}(\mathbf{b}^2,\mu,\zeta)
    &=
    \frac{1}{2}\biggl[\Gamma_\text{cusp}\ln\biggl(\frac{\mu^2}{\zeta}\biggr)-\gamma_V\biggr]
    J_{11}(\mathbf{b}^2,\mu,\zeta)\,.
\end{align}
Here, $\Gamma_\text{cusp}$ is the cusp anomalous dimension and $\gamma_V$ is the anomalous dimension of the quark vector form factor, which to one-loop are given by,
\begin{align}
    \Gamma_\text{cusp}=4a_s C_F+\mathcal{O}(a_s^2)\,,\qquad
    \gamma_V=-6a_sC_F+\mathcal{O}(a_s^2)\,.
\end{align}
The same equations apply to $J^\prime_{11}$ as the renormalization and evolution kernels do not depend on $b$. 

The renormalization and evolution of the twist-3 operators is well known~\cite{Braun:2009mi,Freedman:2014uta,Goerke:2017lei,Beneke:2017ztn,Rodini:2022wki} and involves a convolution in the momentum fraction $\xi$. A few simplifications allow us to recast the evolution presented in refs.~\cite{Braun:2009mi,Rodini:2022wki} in a more elegant form. Since we consider large radii jets, our twist-3 jet functions depend on a single momentum fraction $\xi$. In our case $\xi$ is restricted to the interval $[0,1]$, which can be shown by inserting a complete set of states in between each field that enters the matrix-element definition of the jet function. Because of this constraint, the mixing between different regions of momentum fraction space and mixing between T-even and T-odd contributions, as discussed in ref.~\cite{Rodini:2022wki}, is absent in our case. Following this, the renormalized twist-3 jet functions can be expressed as
\begin{align}
    J_{21,\bar{n}}(\xi,\mathbf{b}^2,\mu,\zeta)
    &=
    \int_0^1\df\xi'\,Z_{J_{21}}(\xi,\xi',\mu,\zeta,\epsilon)\,
    J^\text{bare,sub}_{21,\bar{n}}(\xi',\mathbf{b}^2,\zeta,\epsilon)\,,
\end{align}
and the evolution reads
\begin{align}
    \frac{\df}{\df\ln\mu^2}
    J_{21,\bar{n}}(\xi,\mathbf{b}^2,\mu,\zeta,\delta_E^+)
    &=
    \int_0^1\df\xi'\,\gamma_{J_{21}}(\xi,\xi',\mu,\zeta)\,
    J_{21,\bar{n}}(\xi',\mathbf{b}^2,\mu,\zeta,\delta_E^+)\,.
\end{align}
The anomalous dimension for the twist-3 jet functions has been calculated to one-loop order~\cite{Braun:2009mi} and is given by
\begin{align} \label{eq:ga_J_21}
    \gamma_{J_{21}}(\xi,\xi',\mu,\zeta)
    &=
    a_s\biggl\{\delta(\xi-\xi')
    \biggl[2C_F\ln\biggl(\frac{\mu^2}{\zeta}\biggr)
    +3C_F-2C_A\ln\xi - 2C_F \ln\bar\xi\biggr]
    \nn \\
    &\quad
    +C_A\biggl(\biggl[\frac{\Theta(\xi'\!-\!\xi)}{\xi'\!-\!\xi}\biggr]_+
    \!\!-\! \biggl[\frac{\Theta(\xi\!-\!\xi')}{\xi'\!-\!\xi}\biggr]_+\!
    \!\!+\! \Theta(\xi'\!-\!\xi)\frac{\xi\xi'\!-\!\bar\xi}{\xi'}
    \!-\! \Theta(\xi\!-\!\xi')\frac{\bar\xi(\bar\xi'\!-\!\xi\xi')\!+\!\xi}{\xi \bar\xi'}
    \biggr)
    \nonumber
    \\
    &\quad
    -(2C_F-C_A)\biggl(\Theta(\xi\!-\!\bar{\xi}')\frac{\bar\xi[2\xi-(1+\xi)\bar\xi']}{\xi\xi'}
    +\Theta(\bar\xi'\!-\!\xi)\frac{\xi\xi'}{\bar\xi'}\biggr)
    \biggr\}\,.    
\end{align}

The subtracted and renormalized jet functions also depend on the rapidity scale $\zeta$, in addition to their renormalization scale dependence. For all jet functions the rapidity scale dependence is described by the Collins-Soper kernel, but for $J^\prime_{11}$ there is an extra additive term as the Collins-Soper kernel depends also on $b$. The rapidity evolution is given by
\begin{align}
    \frac{\df}{\df\ln\zeta}J_{11}(\mathbf{b}^2,\mu,\zeta)
    &=
    -\mathcal{D}(\mathbf{b}^2,\mu)\,J_{11}(\mathbf{b}^2,\mu,\zeta)\,,
    \\
    \frac{\df}{\df\ln\zeta}J^\prime_{11}(\mathbf{b}^2,\mu,\zeta)
    &=
    -\mathcal{D}(\mathbf{b}^2,\mu)\,J^\prime_{11}(\mathbf{b}^2,\mu,\zeta)
    -2
    \frac{\df\mathcal{D}(\mathbf{b}^2,\mu)}{\df\ln\mathbf{b}^2}\,
    J_{11}(\mathbf{b}^2,\mu,\zeta)\,,
    \\
    \frac{\df}{\df\ln\zeta}J_{21}(\xi,\mathbf{b}^2,\mu,\zeta)
    &=
    -\mathcal{D}(\mathbf{b}^2,\mu)\,J_{21}(\xi,\mathbf{b}^2,\mu,\zeta)\,,
\end{align}
where $\mathcal{D}(\mathbf{b}^2,\mu)$ is the Collins-Soper kernel.

In principle, one could worry about the role of the subtraction term in the evolution. Specifically, one could argue that the introduction of subtraction terms alters the anomalous dimension. However, the contribution of the subtraction term cancels between the two jet functions in the cross section. Therefore, as long as one evolves the jet functions from the two collinear sectors together, the presence of the subtraction term will not have an effect on phenomenology.

\section{Ingredients at order $a_s$}
\label{sec:JetFunction}

In the previous sections we obtained a factorization formula for the hadronic tensor in terms of a set of jet functions. While the factorization relies on a recoil-free jet axis, the form of the factorization and the above definitions are independent of the precise choice. In order to make a theoretical prediction for the cross section, however, the jet functions must be calculated for a specific jet axis. We discuss the jet axes we consider in \sec{jetaxes}, and we will present the calculation of the corresponding jet functions in \sec{calculation}. 

\subsection{Jet algorithms}
\label{sec:jetaxes}

In this work we consider the jet axes defined through the $E^n$-recombination schemes, with the special case of the winner-take-all (WTA) axis corresponding to the $n \to \infty$ limit. This is a direct generalization of the $p_T^n$-scheme~\cite{Butterworth:2002xg,Cacciari:2011ma} to $e^+e^-$ collisions.
In principle, these axes also depend on the clustering, e.g.~anti-$k_T$ vs.~Cambridge/Aachen, but this distinction is not relevant at the order we are working, where the final state in the fixed-order jet function contains at most two partons\footnote{This dependence has been studied in ref.~\cite{Gutierrez-Reyes:2019vbx} at order $a_s^2 $
using EVENT2~\cite{Catani:1996vz}.}. The resummed jet function will of course contain the dominant effect of multiple emissions.

For all axes, we consider the large radius limit, such that all final-state particles will be clustered into the jet. For this is the reason, we write the final state of the jet function as $\ket{J_\text{alg}}$. The jet axis enters our calculation as follows 
\begin{align}
    \ket{J_\text{alg}}\bra{J_\text{alg}}
    =
    (2\pi)^{d-1}\,\sum_X     
    \delta^{(d-2)}\bigl(\mathbf{P}_J^{\rm alg}\bigr)
    \ket{X}\bra{X}\,.
\end{align}
Here $X$ is the collinear final-state produced by the fields, and the prefactor cancels that of the phase-space integral over the total jet momentum.
Our transverse momentum measurement is encoded in the transverse position of the fields, by fixing our coordinate system such that the jet transverse momentum $\mathbf{P}_J^{\rm alg}$ vanishes. This allowed us to use the same operators as in other transverse momentum measurements.

We now present the specific formulae to determine the transverse momentum of the jet for the $E^n$ scheme, which depends on the number of particles in the final state. In the case where the final state consists of a single parton with momentum $k$, we simply have $\mathbf{P}_J=\mathbf{k}$. When the final state contains two partons with on-shell momenta $k_1$ and $k_2$, respectively, the 3-momentum of the jet in the $E^n$-scheme is given by 
\begin{align} 
    \vec P^{E^n}_J
 &= \frac{k_1^0 + k_2^0}{\sqrt{(k_1^0)^{2n} + 2 (k_1^0)^{n-1} (k_2^0)^{n-1} {\vec k}_1 \sdt {\vec k}_2 + (k_2^0)^{2n}}}\, 
 \bigl[(k_1^0)^{n-1} {\vec k}_1 + (k_2^0)^{n-1} {\vec k}_2\bigr]
\,,\end{align}
where the complicated prefactor arises from the condition that the resulting 4-momentum is massless. However, up to NLP we can approximate ${\vec k}_1 \sdt {\vec k}_2 \approx k_1^0 k_2^0$, since their angle is small due to being in the same collinear direction. This leads to
 \begin{align} \label{eq:rec_scheme}
  \mathbf{P}^{E^n}_J
 = \frac{k_1^0 + k_2^0}{(k_1^0)^n + (k_2^0)^n}\, 
 \bigl[(k_1^0)^{n-1} \mathbf{k}_1 + (k_2^0)^{n-1} {\bf k}_2\bigr] + \mathcal{O}\Bigl(\frac{Q_T^3}{Q^2}\Bigr)
\,.\end{align}
We will restrict ourselves to $n>1$ for which the resulting jet axis is recoil-free and our factorization analysis is justified.  For $n=1$, $\mathbf{P}^{E^n}_J$ is simply the total transverse momentum of particles in the jet, and thus not recoil free.
The WTA recombination scheme corresponds to the $n \to \infty$ limit, in which case \eq{rec_scheme} simplifies to
\begin{align}
    \mathbf{P}^\text{WTA}_J
    &=
    \Theta(k_1^0 - k_2^0)\,\frac{k_1^0 + k_2^0}{k_1^0}\,\mathbf{k}_1
    +\Theta(k_2^0 - k_1^0)\,\frac{k_1^0 + k_2^0}{k_2^0}\, \mathbf{k}_2
\,.\end{align}

\subsection{Calculation}
\label{sec:calculation}

We now calculate all jet functions to first order in $a_s$. For convenience, we summarize their definitions here. The bare and unsubtracted jet functions are defined as
\begin{align} \label{eq:jet_repeat}
    J_{11,\bar{n}}^{\rm bare}(\mathbf{b}^2)
    &=
    \frac{1}{2N_c}\,
    \int\frac{\df y^-}{2\pi}\,e^{\img y^- q^+}\,
    \tr\Bigl[\gamma^+\bra{0}U_{1,\bar{n}}(y^-,b)
    \ket{J^{\bar{n}}_\text{alg}}
    \bra{J^{\bar{n}}_\text{alg}}
    \overline{U}_{1,\bar{n}}(0,0)
    \ket{0}\Bigr]\,,
    \nn \\[1ex]
    J^{\prime\,\text{bare}}_{11,\bar{n}}(\mathbf{b}^2)
    &=
    b^\mu\frac{\partial}{\partial b^\mu} J_{11,\bar{n}}^{\rm bare}(b)\,,
    \nn \\[1ex]
    J_{21,\bar{n}}^{\rm bare}(\xi,\mathbf{b}^2)
    &=
    n^\mu \bar{n}^\nu b^\rho\epsilon_{\mu\nu\rho\sigma}\,
    \frac{1}{\xi - \img\delta_E^+}\,
    \frac{1}{2N_c}    
    \int\frac{\df y_1^-}{2\pi}\frac{\df y_2^-}{2\pi}\,
    e^{\img  (\bar{\xi} y_1^- + \xi y_2^-) q^+}\,
    \nn \\
    &\qquad\times
    \tr\Bigl[\sigma^{\sigma+} \gamma^5
    \bra{0}
    U_{2,\bar{n}}(\{y_1^-,y_2^-\},b)
    \ket{J^{\bar{n}}_\text{alg}}
    \bra{J^{\bar{n}}_\text{alg}}
    \overline{U}_{1,\bar{n}}(0,0)
    \ket{0}\Bigr]\,,
\end{align}
with $U_1$ and $U_2$ defined in \eqs{U1}{U2} and similar definitions hold for the jet functions in the $n$ direction. The subtracted jet functions are given by
\begin{align}
    J^\text{bare,sub}_{11,\bar{n}}(\mathbf{b}^2,\zeta,\epsilon)
    &=
    \frac{J^\text{bare}_{11,\bar{n}} (\mathbf{b}^2,\epsilon,\delta_E^+)}
    {\sqrt{\mathcal{S}_\text{LP}(\mathbf{b}^2,\epsilon,(\delta_E^+)^2\zeta)}}\,,
    \nn\\
    J^{\prime\,\text{bare,sub}}_{11,\bar{n}}
    (\mathbf{b}^2,\zeta,\epsilon)
    &=
    b^\mu\frac{\partial}{\partial b^\mu}
    J^\text{bare,sub}_{11,\bar{n}}(\mathbf{b}^2,\zeta,\epsilon)\,,
    \nn\\
    J^\text{bare,sub}_{21,\bar{n}}(\xi,\mathbf{b}^2,\zeta,\eps)
    &=
    \frac{J^\text{bare}_{21,\bar{n}}(\xi,\mathbf{b}^2)}
        {\sqrt{\mathcal{S}_\text{LP}
    (\mathbf{b}^2,\epsilon,(\delta_E^+)^2\zeta)}}
    -
    S^\text{bare}_{21,\bar{n}}(\xi,\mathbf{b}^2,\epsilon,\delta_E^+,(\delta_E^+)^2\zeta)\,
    J^\text{bare,sub}_{11,\bar{n}}(\mathbf{b}^2,\zeta,\eps)\,,
\end{align}
with 
\begin{align}
    \mathcal{S}_\text{LP}^{\rm bare} 
    (\mathbf{b}^2,\epsilon,2\delta^+\delta^-)
    &=
    \frac{1}{N_c}\tr_c\bigl\{
    \bra{0}[\infty \bar{n} + b,b][b,\infty n + b]
    [\infty n,0][0,\infty \bar{n}]\ket{0}
    \bigr\}\,,
    \\
    S_{21,\bar{n}}^\text{bare}(\xi,\mathbf{b}^2,\epsilon,\delta_E^+,2\delta^+\delta^-)
    &=
    +\frac{b_\rho}{\xi-\img \delta_E^+}
    \int\frac{\df y_2^-}{2\pi}\,e^{i\xi y_2^- q^+}\,
    \frac{1}{N_c}\tr\Bigl[\bra{0}
    [b + y_2^- n + \infty n,b + y_2^- n]
    \nn \\ & \quad \times
    [b + y_2^- n,b + y_2^- n + \infty\bar{n}]
    gF^{\rho+}(y_2^- n + b)\,
    \ket{X}\bra{X}
    [\infty \bar{n},0]
    [0,\infty n]
    \ket{0}\Bigr]\,.
\nn \end{align}
Finally, the renormalized jet functions are given by
\begin{align}
    J_{11,\bar{n}}(\mathbf{b}^2,\mu,\zeta)
    &=
    Z_{J_{11}}(\zeta,\mu,\epsilon)\,
    J^\text{bare,sub}_{11,\bar{n}}
    (\mathbf{b}^2,\zeta,\epsilon)\,,
    \nn\\
    J_{21,\bar{n}}(\xi,\mathbf{b}^2,\mu,\zeta)
    &=
    \int_0^1\df\xi'\,Z_{J_{21}}(\xi,\xi',\mu,\zeta,\epsilon)\,
    J^\text{bare,sub}_{21,\bar{n}}(\xi',\mathbf{b}^2,\zeta,\epsilon)\,.
\end{align}

\subsubsection*{Bare jet functions}

\begin{figure}
\centering
\begin{fmffile}{diagrams}
\fmfcmd{
    path quadrant, q[], otimes;
    quadrant = (0, 0) -- (0.5, 0) & quartercircle & (0, 0.5) -- (0, 0);
    for i=1 upto 4: q[i] = quadrant rotated (45 + 90*i); endfor
    otimes = q[1] & q[2] & q[3] & q[4] -- cycle;
}
\fmfwizard
\begin{fmfgraph*}(110,70)
\thirteenbyninegrid
\fmf{dashes,tension=0}{g1,g9}
\fmf{fermion,tension=0}{a2,m2}
\fmf{gluon,tension=0,right}{j2,d2}
\fmf{double,tension=0}{m2,m8}
\fmf{double,tension=0}{a2,a8}
\fmfv{d.sh=otimes,d.f=empty,d.si=.08w}{a2}
\fmfv{d.sh=otimes,d.f=empty,d.si=.08w}{m2}
\end{fmfgraph*}
\hspace{7ex}
\begin{fmfgraph*}(110,70)
\thirteenbyninegrid
\fmf{dashes,tension=0}{g1,g9}
\fmf{plain,tension=0}{a2,d2}
\fmf{fermion,tension=0}{d2,m2}
\fmf{double,tension=0}{m2,m8}
\fmf{double,tension=0}{a2,a8}
\fmf{gluon,tension=0}{m5,c2}
\fmfv{d.sh=otimes,d.f=empty,d.si=.08w}{a2}
\fmfv{d.sh=otimes,d.f=empty,d.si=.08w}{m2}
\end{fmfgraph*}
\hspace{7ex}
\begin{fmfgraph*}(110,70)
\thirteenbyninegrid
\fmf{dashes,tension=0}{g1,g9}
\fmf{plain,tension=0}{a2,d2}
\fmf{fermion,tension=0}{d2,m2}
\fmf{double,tension=0}{m2,m8}
\fmf{double,tension=0}{a2,a8}
\fmf{gluon,tension=0}{m5,c2}
\fmfv{d.sh=otimes,d.f=empty,d.si=.08w}{a2}
\fmfv{d.sh=otimes,d.f=empty,d.si=.08w}{m2}
\fmfv{d.sh=otimes,d.f=empty,d.si=.08w}{m5}
\end{fmfgraph*}
\end{fmffile}
\caption{\label{fig:diagrams} Diagrams for the matrix elements that enter the calculation of the twist-2 jet function at NLO (left and middle) and the twist-3 jet function at LO (right). The double line represents the Wilson line, and the $\otimes$ denote insertions of the quark field (at the corners) and gluon field strength (on the double line). The diagrams that give a vanishing result are not shown. }
\end{figure}
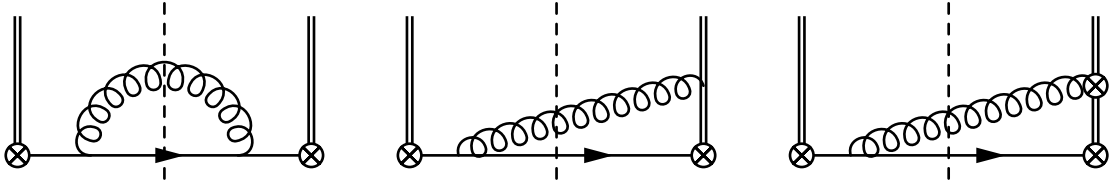

The (leading-order) matrix elements that are required in this calculation are given by
\begin{align} \label{eq:me}
    \bra{0} U_{1,\bar{n}}(y^-,b) \ket{q(p)}^{(0)}
    &=
    e^{\img \mathbf{b}\cdot\mathbf{p}}
    e^{-\img y^- p^+}\,
    u(p)\,,
    \\[1ex]
    \bra{0} U_{1,\bar{n}}(y^-,b) \ket{q(p-k) g(k)}^{(0)}
    &=
    e^{\img \mathbf{b}\cdot\mathbf{p}}
    e^{-\img y^- p^+}\,
    (-g)
    \biggl[\frac{\slashed{p}\gamma^\nu}{p^2} 
        -\frac{n^\nu}{k^+ - \img \delta^+}\biggr]
    t^a\,u(p-k) \epsilon^a_\nu(k)\,,
    \nonumber
    \\[1ex]
    \bra{0} U_{2,\bar{n}}(\{y_1^-,y_2^-\},b) \ket{q(p\!-\!k) g(k)}^{(0)}
    &=
    e^{\img \mathbf{b}\cdot\mathbf{p}}
    e^{-\img  y_1^- (p-k)^+}
    e^{-\img  y_2^- k^+}
    (-\img g)
    \bigl[\slashed{k}_T n^\mu \!-\! k^+ \gamma_T^\mu\bigr] 
    t^a\,u(p\!-\!k) \epsilon_\mu^a(k)\,,
\nn\end{align}
with the relevant diagrams shown in \fig{diagrams}. The index $a$ on the polarization vector indicates the color of the gluon (though technically is not part of the polarization).
We have omitted the virtual corrections, as the virtual correction to $J_{11}$ vanishes in dimensional regularization, and the virtual correction to $J_{21}$ is exactly zero. Inserting the expressions in \eq{me} in \eq{jet_repeat}, we find to order $a_s$ 
\begin{align}
    J_{11,\bar{n}}^{\rm bare}
    &=
    1
    + \ 
    4g^2 C_F\, 
    \int\frac{\df^d p}{(2\pi)^d}\frac{\df^d k}{(2\pi)^d}\,
    (2\pi)\delta_+(k^2)\,(2\pi)\delta_+[(p-k)^2]\,
    (2\pi)^{d-1}\delta^{(d-2)}(\mathbf{P}_J^{\rm alg})
    \nonumber
    \\
    &\qquad\times
    e^{\img \mathbf{b}\cdot\mathbf{p}}\,
    \delta(p^+ - q^+)\,
    \frac{k^+}{p^2}
    \biggl[\frac{(p^+)^2+(p^+-k^+)^2}{(k^+)^2 + (\delta^+)^2}
        -\epsilon\biggr]\,,
    \\[2ex]
    J_{21,\bar{n}}^{\rm bare}
    &=
    -4\img g^2C_F\,\frac{1}{\xi - \img\delta_E^+}
    \int\frac{\df^d p}{(2\pi)^d}\frac{\df^d k}{(2\pi)^d}\,
    (2\pi)\delta_+(k^2)\,(2\pi)\delta_+[(p-k)^2]\,
    (2\pi)^{d-1}\delta^{(d-2)}(\mathbf{P}_J^{\rm alg})
    \nonumber
    \\
    &\qquad\times
    e^{\img \mathbf{b}\cdot\mathbf{p}}\,
    \delta(q^+ - p^+)
    \delta(\xi q^+ - k^+)\,
    \frac{(q^+)^2}{p^2}\,
    (\bar\xi+\epsilon\xi)
    \bigl[\bar\xi k^\mu - \xi (p-k)^\mu\bigr]b_\mu\,.
\end{align}
These expressions only depend on the details of the jet algorithm through $\mathbf{P}_J^{\rm alg}$.

From here, one performs the integral over $\mathbf{k}$ using $\delta^{(d-2)}(\mathbf{P}_J^{\rm alg})$, and the lightcone components of $p$ and $k$. For the $E^n$ scheme, 
\begin{align}
    p^2 = \frac{\xi\bar\xi(\bar\xi^n+\xi^n)^2}{(\xi^n\bar\xi-\xi\bar{\xi}^n)^2} \mathbf{p}^2
\,,\end{align}
leading to
\begin{align}
    J^\text{bare}_{11,\bar{n}}
    &=
    1
    +
    \frac{g^2 C_F}{(2\pi)^{d-1}}
    \biggl(\int\df^{d-2}{\bf p}\,
    \frac{e^{\img \mathbf{b}\cdot\mathbf{p}}}{\mathbf{p}^2}\biggr)
    \int_0^1\df\xi\,
    \biggl|\frac{\bar\xi^n + \xi^n}{\xi^{n-1}-\bar\xi^{n-1}}\biggr|^{d-4}
    \biggl[\frac{\xi(1+\bar\xi^2)}{\xi^2 + (\delta_E^+)^2}
        -\epsilon\xi\biggr]\,,
    \\[2ex]
    J^\text{bare}_{21,\bar{n}}
    &=
    -\frac{\img g^2C_F}{(2\pi)^{d-1}}
    \biggl(\int\df^{2}{\bf p}\,
    \frac{e^{\img \mathbf{b}\cdot\mathbf{p}}}{\mathbf{p}^2}\, \mathbf{p}\cdot\mathbf{b}\biggr)
    \Theta(\xi)\Theta(\bar\xi)
    \frac{\bar\xi}{\xi-\img\delta_E^+}\,
    \frac{\xi^n + \bar\xi^n}{\xi^{n-1}-\bar\xi^{n-1}}
    \,.
\end{align}
The remaining integral over $\mathbf{p}$ in $J_{11}$ can then be performed by using 
\begin{align}\label{eq:pTintegral}
    \int\df^{d-2}\mathbf{p}\,
    \frac{e^{\img \mathbf{b}\cdot\mathbf{p}}}{\mathbf{p}^2}
    = 4^{-\eps} \pi^{1-\eps} \Gamma(-\eps)\,
    (\mathbf{b}^2)^\epsilon\,,    
\end{align}
and the integral that appears for $J_{21}$ can be obtained by differentiating with respect to $\mathbf{b}$. Finally, we perform the integral over $\xi$ for $J_{11,\bar{n}}$ and expand the result in $\epsilon$
\begin{align}
    J^\text{bare}_{11,\bar{n}}
    &=
    1
    +
    a_s C_F
    \biggl\{
    \biggl[\frac{1}{\epsilon}+
    \ln\Bigl(\frac{\mu_0^2\mathbf{b}^2}{4e^{-2\gamma_E}}\Bigr)\biggr]
    (3+4\ln\delta_E^+)
    +J_{C_F}
    \biggr\}
    \,,
    \\[2ex]
    J^\text{bare}_{21,\bar{n}}
    &=
    -4a_s C_F\,
    \Theta(\xi)\Theta(\bar\xi)
    \frac{\bar\xi}{\xi-\img\delta_E^+}\,
    \frac{\xi^n + \bar\xi^n}{\bar\xi^{n-1}-\xi^{n-1}}
    \,,
\end{align}
with
\begin{align}
  J_{C_F} = 1+ 4 \int_0^1\!\df\xi\,\frac{1+(1-\xi)^2}{\xi}\, 
  \ln\biggl|\frac{(1-\xi)^n + \xi^n}{(1-\xi)^{n-1}-\xi^{n-1}}\biggr|
\,.\end{align}

\subsubsection*{Subtraction and renormalization}

To obtain the subtracted and renormalized jet functions at first order in $a_s$, we need the LP and NLP soft functions and the renormalization factor $Z_{J_{11}}$ to order $a_s$. We do not need the one-loop expression for $Z_{J_{21}}$ at this order. The LP soft function with $\delta$ regulator has been calculated to one-loop order in ref.~\cite{Echevarria:2011epo}, and the result reads
\begin{align}
    \mathcal{S}_\text{LP}(\mathbf{b}^2,\epsilon,2\delta^+\delta^-)
    &=
    1-4a_s C_F
    \biggl\{\frac{1}{\epsilon^2}-
    \biggl[\frac{1}{\epsilon}
            +\ln\biggl(\frac{\mu^2 \mathbf{b}^2}{4 e^{-2\ga_E}}\biggr)\biggr]
        \ln\biggl(\frac{2\delta^+ \delta^-}{\mu^2}\biggr)
        -\frac{1}{2}\ln^2\biggl(\frac{\mu^2 \mathbf{b}^2}{4 e^{-2\ga_E}}\biggr)
        -\frac{\pi^2}{12} 
    \biggr\} 
    \nn \\ & \quad
    +\mathcal{O}(a_s^2)\,.
\end{align}
The renormalization factor for $J_{11}$ is given by
\begin{align}
    Z_{J_{11}}(\zeta,\mu,\eps) = 1 + 2a_s  C_F
    \biggl\{-\frac{1}{\epsilon^2}
        -\frac{1}{\epsilon}\biggl[\ln\Bigl(\frac{\mu^2}{\zeta}\Bigr)
            +\frac{3}{2}\biggr]\biggr\}
\,.\end{align}

\begin{figure}
\centering
\begin{fmffile}{NLPsoft}
\fmfcmd{
    path quadrant, q[], otimes;
    quadrant = (0, 0) -- (0.5, 0) & quartercircle & (0, 0.5) -- (0, 0);
    for i=1 upto 4: q[i] = quadrant rotated (45 + 90*i); endfor
    otimes = q[1] & q[2] & q[3] & q[4] -- cycle;
}
\fmfwizard
\begin{fmfgraph*}(120,100)
\thirteenbyninegrid
\fmf{dashes,tension=0}{g1,g9}
\fmf{gluon,tension=0}{c7,m5}
\fmf{double,tension=0}{a5,e9}
\fmf{double,tension=0}{a5,e1}
\fmf{double,tension=0}{m5,i9}
\fmf{double,tension=0}{m5,i1}
\fmfv{d.sh=otimes,d.f=empty,d.si=.1w}{m5}
\end{fmfgraph*}
\end{fmffile}
\caption{\label{fig:NLPsoft} The only diagram contributing to the NLP soft function at leading order. The $\otimes$ denotes the field strength operator and the double lines represent the Wilson lines.}
\end{figure}
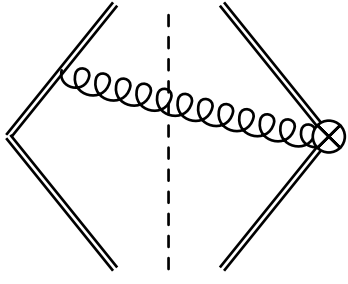

The NLP soft function can be calculated from its matrix element definition. At leading order only one diagram contributes, depicted in \fig{NLPsoft}. The diagrams that do not contribute are two virtual diagrams which vanish in dimensional regularization and a real-emission diagram that vanishes because $n^2=\bar{n}^2=0$. The remaining diagram results in
\begin{align}
    S_{21,\bar{n}}^\text{bare}
    (\xi,\mathbf{b}^2,\epsilon,\delta_E^+,2\delta^+\delta^-)
    &=
    \frac{g^2 C_F}{(2\pi)^{d-1}}
    \frac{b_\rho\partial^\rho}{\xi-\img \delta_E^+}
    \int\df^d k\,\delta_+(k^2)\delta(\xi q^+ - k^+)\,
    \frac{e^{-\img b\cdot k}}{k^- - \img\delta^-}
    +\mathcal{O}(a_s^2)\,.
\end{align}
The integrals over $k^+$ and $k^-$ can be performed using the delta functions and the remaining integral over $\mathbf{k}$ can be performed using \eq{pTintegral}. The final result for the NLP soft function reads
\begin{align}
    S_{21,\bar{n}}^\text{bare}
    (\xi,\mathbf{b}^2,\epsilon,\delta_E^+,2\delta^+\delta^-)
    &=
    -4a_s C_F\,\Theta(\xi)\frac{1}{\xi-\img \delta_E^+}
    +\mathcal{O}(a_s^2)\,.
\end{align}

\subsubsection*{Final results}

\begin{figure}
{\centering
\includegraphics[width=0.6\textwidth]{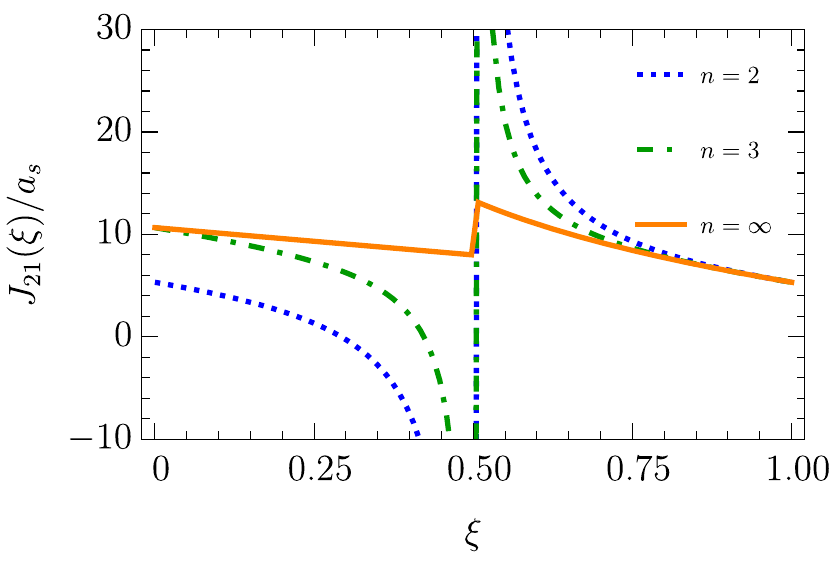}\\[-2ex]}
\caption{\label{fig:J21} The twist-3 jet function $J_{21}$ in \eq{J_res}, divided by $a_s$, as function of $x$ for  $n = 2, 3, \infty$.}
\end{figure}

We now present the final results for all the ingredients of the cross section to first order in $a_s$. The subtracted and renormalized jet functions read 
\begin{align} \label{eq:J_res}
J^{E^n}_{11,\bar{n}}(\mathbf{b}^2)
    &=
    1
    +
    a_s C_F\biggl\{-\ln^2\biggl(\frac{\mu^2 \mathbf{b}^2}{4 e^{-2\ga_E}}\biggr) + \ln\biggl(\frac{\mu^2 \mathbf{b}^2}{4 e^{-2\ga_E}}\biggr) \biggl[3 + 2 \ln \Bigl(\frac{\mu^2}{\zeta}\Bigr)\biggr] + J_{C_F} - \frac{\pi^2}{6}\biggr\}\,,
    \nn\\
    J^{\prime\,E^n}_{11,\bar{n}}(\mathbf{b}^2)
    &=
    2a_s C_F\biggl\{-2\ln\biggl(\frac{\zeta \mathbf{b}^2}{4 e^{-2\ga_E}}\biggr) +3\biggr\}\,,
    \nn\\
    J^{E^n}_{21,\bar{n}}(\xi,\mathbf{b}^2)
    &=
    -4a_s C_F\,\Theta(\xi)\Theta(\bar\xi)\,
    \frac{[1+\xi-\xi^2] \xi^{n-2} - [2-\xi] (1-\xi)^{n-1}}
        {(1-\xi)^{n-1}-\xi^{n-1}}\,.
\end{align}
Note that since $J_{21}$ and $J^\prime_{11}$ only start at order $a_s$, the renormalization does not effect them at this order in perturbation theory. The WTA axis corresponds to the $n \to \infty$ limit. In this case, 
\begin{align}
   \lim_{n \to \infty} J_{C_F} &=
    7-\frac{2\pi^2}{3}-6\log (2)
   \,,\nn \\
   \lim_{n \to \infty} 
   \frac{[1+\xi-\xi^2] \xi^{n-2} - [2-\xi](1-\xi)^{n-1}}
        {(1-\xi)^{n-1}-\xi^{n-1}}
    &=
    -(2-\xi) \Theta(\tfrac{1}{2}-\xi)
    -\frac{1+\xi-\xi^2}{\xi}\Theta(\xi-\tfrac{1}{2})\,.
\end{align}
The twist-3 jet function $J_{21}$ is shown in \fig{J21} for $n=2,3$ and $\infty$ (which corresponds to WTA). By dividing $J_{21}$ by $a_s$, we removed dependence on the renormalization scale. Note that these jet functions have a singularity at $\xi=\tfrac12$ arising from the $E^n$ scheme. However, since the hard function is smooth at this point, 
there is a cancellation between $\xi=\tfrac12 + \de \xi$ and $\xi=\tfrac12 - \de \xi$ to yield a finite result (principle value). For completeness, we also present the one-loop results for the hard functions,
\begin{align}
    H_1(Q^2,\mu)
    &=
    1+2a_s C_F\biggl(
    -\ln^2\Bigl(\frac{Q^2}{\mu^2}\Bigr) 
    +3\ln\Bigl(\frac{Q^2}{\mu^2}\Bigr)
    -8+\frac{7\pi^2}{6}\biggr)\,,
    \nn\\
    H_2(\xi,Q^2,\mu)
    &=
    1+2a_s\biggl[
    C_F\biggl(-\ln^2\Bigl(\frac{Q^2}{\mu^2}\Bigr)
        +2\ln\Bigl(\frac{Q^2}{\mu^2}\Bigr)
        -\frac{11}{2}+\frac{7\pi^2}{6}-\img\pi\biggr)
    -\tfrac{1}{2}C_A\frac{\ln\xi}{1-\xi}
    \nn\\
    &\quad\qquad\quad
    -\bigl(C_F-\tfrac{1}{2}C_A\bigr)
        \biggl(\ln\Bigl(\frac{Q^2}{\mu^2}\Bigr) 
        +\img\pi+\tfrac{1}{2}\ln(1-\xi)-2\biggr)
        \frac{\ln(1-\xi)}{\xi}\biggr]\,.
\end{align}

\section{Conclusions}
\label{sec:Conclusions}

Due to the very high precision of the Large Hadron Collider and the advent of the Electron-Ion Colliders, there has been an upsurge in interest surrounding TMDs, as this presents a unique opportunity to determine these distributions with unprecedented precision. This necessitates a deeper understanding of the factorization theorem for transverse momentum dependent cross sections. 

In this study, we examine select aspects of the recently developed operator basis that contribute to the differential cross sections at next-to-leading power. We focus on jets,  yielding an infrared and collinear safe observable, that can be calculated in perturbation theory. In order to further simplify the description, we consider only large radii such that all energetic radiation is inside jets. By using a recoil-free jet recombination schemes, we avoid non-global logarithms.

It is natural to examine how higher twist operators that appear at NLP enter in the description of jets and generate new measurable effects. We illustrate all these aspects on a prototypical process, $e^+e^-\rightarrow 2$ jets, which is theoretically simple and experimentally relevant. The final factorized cross section in \eq{sigmabare} was obtained by starting from the power expansion of the hadronic tensor, decomposing in spin structures and applying discrete symmetries. It involves only one independent twist-2 and twist-3 jet operator matrix element for each collinear direction.  To our knowledge, this is the first time that the contribution of a twist-3 operator to a jet has been determined. We have found that NLP effects exist in $e^+e^-\rightarrow 2$ jets only when the two jets are different, e.g.~when two different recoil-free jet algorithms are used or a jet is replaced by an identified hadron (TMD fragmentation function). 

It is well known that factorization theorems at sub-leading power contain additional divergences that require careful treatment. In sec.~\ref{sec:special} we have shown that these divergences can be subtracted by means of a soft function and cancel in the cross section between the two collinear sectors. Moreover, we demonstrated that the cancellation of these divergences requires that the jet functions of the two collinear sectors are evaluated at the same rapidity and renormalization scales, and  they need to be evolved together when considering resummation. The evolution equations necessary to perform resummation were presented in sec.~\ref{sec:evolution}.

Having arrived at a finite and factorized expression for the cross section, we computed the twist-3 jet functions using WTA and E$^n$ recombination schemes at leading order in perturbation theory. We discussed how these jet algorithms can be implemented in perturbative calculations. All results that are needed to obtain a resummed expression for the cross section are presented to order  $a_s$ in sec.~\ref{sec:calculation}.

There are several aspects of this study that we consider relevant for future applications. First of all, the measurement we consider allows for a substantial simplification of the NLP formalism, making it more palatable. The process that we describe can be measured at the Belle experiment, but we leave a detailed phenomenological analysis to future work. It would be intriguing to consider how the nonperturbative effects that we study impact QCD research at the Future Circular Collider (FCC). More importantly, measurements of hadron structure at the EIC will benefit from our NLP jet analysis, which we consider a priority. Our framework can be naturally extended to also take into account polarized targets at the EIC.

\acknowledgments

We thank Lorenzo Zoppi for collaborating in early stages of this project, and Anjie Gao, Johannes Michel and  Alexey Vladimirov for discussions on TMDs at next-to-leading power. This project is supported by the Spanish Ministry grant PID2019-106080GB-C21 and PID2022-136510NB-C31, European Union Horizon 2020 research and innovation program under grant agreement num.~824093 (STRONG-2020), the NWO projectruimte 680-91-122, and the D-ITP consortium, a program of NWO that is funded by the Dutch Ministry of Education, Culture and Science (OCW).

\appendix
\section{Fierz relations}
\label{sec:Fierz}

The jet functions defined in \eqs{J_11_indices}{J_21_indices} have open spinor indices. These spinor indices are at first contracted between the jet functions for the two different collinear directions, see \eq{W_intermediate}. To remove contractions between different jet functions and obtain our factorized hadronic tensor, we apply the following Fierz relations
\begin{align} \label{eq:Fierz_full}
    4(\gamma^\mu)_{ij}(\gamma^\nu)_{kl}
    &=
    g^{\mu\nu}
    \Bigl[
    \delta_{il} \delta_{kj}
    +(\img\gamma^5)_{il} (\img\gamma^5)_{kj}
    -(\gamma^\alpha)_{il} (\gamma_\alpha)_{kj}
    \nonumber
    \\
    &\quad\qquad\quad
    -(\gamma^\alpha \gamma^5)_{il} (\gamma_\alpha \gamma^5)_{kj}
    +\frac{1}{2} (\img\sigma^{\alpha\beta} \gamma^5)_{il}
        (\img\sigma_{\alpha\beta} \gamma^5)_{kj}
    \Bigr]
    \nonumber
    \\
    &\quad
    +(\gamma^{\{\mu})_{il} (\gamma^{\nu\}})_{kj}
    +(\gamma^{\{\mu}\gamma^5)_{il} (\gamma^{\nu\}}\gamma^5)_{kj}
    -(\img\sigma^{\alpha\{\mu} \gamma^5)_{il} 
        (\img{\sigma^{\nu\}}}_\alpha \gamma^5)_{kj}
    \nonumber
    \\
    &\quad
    -\frac{\img}{2}\epsilon^{\mu\nu\lambda\eta}
    \Bigl[
    (\img\sigma^{\lambda\eta} \gamma^5)_{il} \delta_{kj}
    -\delta_{il} (\img\sigma^{\lambda\eta}\gamma^5)_{kj}
    \Bigr]
    \nonumber
    \\
    &\quad
    +(\img\sigma^{\mu\nu} \gamma^5)_{il} (\gamma^5)_{kj}
    +(\gamma^5)_{il}(\img\sigma^{\mu\nu} \gamma^5)_{kj}
    \nonumber
    \\ 
    &\quad
    +\img\epsilon^{\mu\nu\alpha\beta}
    \Bigl[
    (\gamma^\alpha \gamma^5)_{il} (\gamma^\beta)_{kj}
    +(\gamma^\alpha)_{il} (\gamma^\beta \gamma^5)_{kj}
    \Bigr]\,,
    \nonumber
    \\[2ex]
    4 (\gamma^\mu)_{ij} \delta_{kl}
    &=
    (\gamma^\mu)_{il} \delta_{kj}
    +\delta_{il} (\gamma^\mu)_{kj}
    +(\gamma^5)_{il} (\gamma^\mu \gamma^5)_{kj}
    -(\gamma^\mu \gamma^5)_{il}(\gamma^5)_{kj}
    \nonumber
    \\ 
    &\quad
    -\frac{\img}{2}\epsilon^{\mu\nu\tau\eta}
    \Bigl[
    (\img\sigma^{\nu\tau} \gamma^5)_{il} (\gamma^\eta)_{kj}
    -(\gamma^\eta)_{il}(\img\sigma^{\nu\tau}\gamma^5)_{kj}
    \Bigr]
    \nonumber
    \\
    &\quad
    +(\gamma^\eta \gamma^5)_{il} (\img\sigma^{\mu\eta} \gamma^5)_{kj} 
    +(\img\sigma^{\mu\eta} \gamma^5)_{il} (\gamma^\eta \gamma^5)_{kj}\,.
\end{align}

Inserting these expressions in \eq{W_intermediate}, many of the above Dirac structures lead to vanishing expressions. Due to the collinear nature of the jet functions, only structures which contain an explicit $\gamma^+$ or $\gamma^-$ survive (see \eq{good_bad}). Moreover, because the jet algorithm/measurement is spin-independent and due to parity symmetry, we are left with  
\begin{align}
    4(\gamma_T^\mu)_{ij}(\gamma_T^\nu)_{kl}
    &=
    -g_T^{\mu\nu}
    \Bigl[
    (\gamma^+)_{il} (\gamma^-)_{kj}
    +(\gamma^-)_{il} (\gamma^+)_{kj}
    \Bigr] + \dots\,,
    \nn \\[2ex]
    4 (\gamma_T^\mu)_{ij} \delta_{kl}
    &=
    +\img\epsilon_T^{\mu\alpha}
    \Bigl[
    (\gamma^+)_{il}(\img\sigma^{\alpha-}\gamma^5)_{kj}
    +(\img\sigma^{\alpha+} \gamma^5)_{il} (\gamma^-)_{kj}
    \nn \\
    &\quad\qquad\quad
    -(\gamma^-)_{il}(\img\sigma^{\alpha+}\gamma^5)_{kj}
    -(\img\sigma^{\alpha-} \gamma^5)_{il} (\gamma^+)_{kj}
    \Bigr] + \dots\,,
\end{align}
where the ``$\dots$" are terms in \eq{Fierz_full} which give a vanishing contribution when inserted into \eq{W_intermediate}.

\normalbaselines 

\bibliographystyle{JHEP}
\bibliography{bibFILE}

\end{document}